\documentclass[a4paper,fleqn,usenatbib]{mnras}
\usepackage{newtxtext,newtxmath}

\usepackage{ae,aecompl}
\usepackage{graphicx}	
\usepackage{amsmath}	
\usepackage{amssymb}	
\usepackage{mathtools}
\usepackage{hyperref}
\usepackage{lineno}

\title[GAMA: sSFR-M$_{*}$ Part I - $\sigma_{\mathrm{sSFR}}$-M$_{*}$]{Galaxy And Mass Assembly (GAMA): The sSFR-M$_{*}$ relation part I -- $\sigma_{\mathrm{sSFR}}$-M$_{*}$ as a function of sample, SFR indicator and environment}
\author[L. J. M. Davies et. al.]{L. J. M. Davies$^{1}$\thanks{E-mail:
 luke.j.davies@uwa.edu.au}, C. del P. Lagos$^{1,2}$, A. Katsianis$^{3}$, A. S. G. Robotham$^{1,2}$, \newauthor L. Cortese$^{1,2}$,  S. P. Driver$^{1}$,   M. N. Bremer$^{4}$, M. J. I. Brown$^{5}$, S. Brough$^{6}$, \newauthor M. E. Cluver$^{7,8}$,  M.W. Grootes$^{9}$, B. W. Holwerda$^{10}$,  M. Owers$^{11}$, S. Phillipps$^{4}$   \\
\\ 
$^{1}$ ICRAR, The University of Western Australia, 35 Stirling Highway, Crawley, WA 6009, Australia\\
$^{2}$ ARC Centre of Excellence for All Sky Astrophysics in 3 Dimensions (ASTRO 3D)\\
$^{3}$ Department of Astronomy, Universitad de Chile, Camino El Observatorio 1515, Las Condes, Santiago, Chile\\
$^{4}$ School of Physics, University of Bristol, Tyndall Avenue, Bristol BS8 1TL, UK\\
$^{5}$ School of Physics and Astronomy, Monash University, Clayton, Victoria 3800, Australia\\
$^{6}$ School of Physics, University of New South Wales, NSW 2052, Australia\\ 
$^{7}$Centre for Astrophysics and Supercomputing, Swinburne University of Technology, John Street, Hawthorn, 3122, Australia \\
$^{8}$ Department of Physics and Astronomy, University of the Western Cape, Robert Sobukwe Road, Bellville, 7535, South Africa \\
$^{9}$ ESA/ESTEC SCI-S, Keplerlaan 1, 2201 AZ Noordwijk, The Netherlands \\
$^{10}$ Department of Physics and Astronomy, 102 Natural Science Building, University of Louisville, Louisville KY 40292, USA\\
$^{11}$ Department of Physics and Astronomy, Macquarie University, North Ryde, NSW 2109, Australia
}

\date{Accepted XXX. Received YYY; in original form ZZZ}

\pubyear{2018}

\begin{document}
\label{firstpage}
\pagerange{\pageref{firstpage}--\pageref{lastpage}}
\maketitle

\begin{abstract}

Recently a number of studies have proposed that the dispersion along the star formation rate - stellar mass relation ($\sigma_{\mathrm{sSFR}}$-M$_{*}$) is indicative of variations in star-formation history (SFH) driven by feedback processes. They found a `U'-shaped dispersion and attribute the increased scatter at low and high stellar masses to stellar and active galactic nuclei feed-back respectively. However, measuring $\sigma_{\mathrm{sSFR}}$ and the shape of the $\sigma_{\mathrm{sSFR}}$-M$_{*}$ relation is problematic and can vary dramatically depending on the sample selected, chosen separation of passive/star-forming systems, and method of deriving star-formation rates ($i.e.$ H$\alpha$ emission vs spectral energy distribution fitting). As such, any astrophysical conclusions drawn from measurements of $\sigma_{\mathrm{sSFR}}$ must consider these dependencies. Here we use the Galaxy And Mass Assembly survey to explore how $\sigma_{\mathrm{sSFR}}$ varies with SFR indicator for a variety of selections for disc-like `main sequence' star-forming galaxies including colour, star-formation rate, visual morphology, bulge-to-total mass ratio, S\'{e}rsic index and mixture modelling. We find that irrespective of sample selection and/or SFR indicator, the dispersion along the sSFR-M$_{*}$ relation does follow a `U'-shaped distribution. This suggests that the shape is physical and not an artefact of sample selection or method. We then compare the $\sigma_{\mathrm{sSFR}}$-M$_{*}$ relation to state-of-the-art hydrodynamical and semi-analytic models and find good agreement with our observed results. Finally, we find that for group satellites this `U'-shaped distribution is not observed due to additional high scatter populations at intermediate stellar masses.

\end{abstract}


\section{Introduction}

Star-forming galaxies over a range of epochs and environments have been found to display a tight correlation between their star formation rate (SFR) and stellar mass (M$_{*}$), described as the the star-forming sequence \citep[SFS, or star-forming `main sequence', ][]{Elbaz07,Noeske07,Salim07,Whitaker12,Davies16b}. This sequence has been shown to be largely linear out to high redshift, but with increasing normalisation as a function of look-back time \citep[$e.g.$][]{Schreiber15, Lee15}. The physical interpretation of these observations \citep[$i.e.$][]{Bouche10, Daddi10, Genzel10, Lagos11, Lilly13, Dave13, Mitchell16} is that the bulk of star-forming galaxies reside in a self-regulated equilibrium state where the inflow rate of gas for future star-formation is balanced by the rate at which new stars are formed and the outflow of gas from feedback events ($i.e.$ Supernovae, SNe, and Active Galactic Nuclei, AGN). 

However, within the full sSFR-M$_{*}$ plane the situation is more complex. While this simple self-regulated model is likely to fit for sources which sit close to the locus of the SFS, there exist various other populations which deviate from this model, such as the passive cloud which sits below the SFS, `green valley' sources which sit between the SFS and passive cloud, and star-bursting sources which reside above the SFS. More recent results have found that galaxies move significantly within the SFS over their lifetime based on small star-burst/quenching events \citep[$i.e.$][]{Magdis12, Tacchella16}, such that the locus of the SFS remains constant at a given epoch, but with individual galaxies move within the SFS producing the observed scatter. In addition, there has been recent evidence to the suggest that the SFS is non-linear in the high stellar mass regime and flattens \citep[$e.g.$][]{Rodighiero10, Elbaz11, Whitaker12,Lee15,Katsianis16, Grootes17, Grootes18}. This is likely to be caused by the inclusion of non-star-forming bulge components in stellar mass estimates at log$_{10}$[M$_{*}$/M$_{\odot}$]$>$10 \citep[][]{Erfanianfar16}. The flattening is found to be removed when only considering disc-dominated systems \citep[$e.g.$][]{Abramson14,Willett15} and/or just the disc components of galaxies (Davies et al., in preparation - paper II in this series - and Cook et al., in preparation).               

The position of a galaxy within the sSFR-M$_{*}$ plane is largely determined by its star-formation history \citep[SFH,][]{Madau98, Kauffmann03}. This history is governed by many events which occur in the lifetime of the galaxy which can fundamentally affect its trajectory through the sSFR-M$_{*}$ plane, such as gas accretion \citep{Kauffmann06,Sancisi08, Mitchell16}, mergers \citep[$e.g.$][and see review of Conselice 2014]{Bundy04, Baugh06, Kartaltepe07, Bundy09,Jogee09,deRavel09,Lotz11,Robotham14}, SNe- \citep{Dekel86, DallaVecchia08,Scannapieco08} and AGN-feedback \citep{Kauffmann04,Fabian12}, environmental effects such as starvation, strangulation and stripping \citep[$e.g.$][]{Giovanelli85, Moore99, Peng10, Cortese11, Darvish16}, and morphological changes \citep{Conselice14, Eales15}. It is the combination of these SFHs that ultimately result in the distribution of points in the sSFR-M$_{*}$ plane \citep[][]{Abramson16}.  As such, understanding the global distribution of sources within this plane, the position of various sub-populations split on properties such as environment, morphology and structure, and physical mechanisms which result in galaxies moving through the plane is essential to our parameterisation of the factors driving galaxy evolution. 

One key diagnostic of these physical mechanisms is the dispersion along the sSFR-M$_{*}$ relation \citep[$\sigma_{\mathrm{sSFR}}$-M$_{*}$,][Katsianis et al in preparation]{Guo15, Willett15}. This dispersion is essentially a metric of the variation in a galaxy's recent SFH at a given stellar mass. For example, recent quenching and star-burst events push galaxies below and above the SFS respectively, increasing the dispersion. In addition, as these events have opposite effects in terms of sSFR, asymmetry in the distribution of points about the SFS at a given mass is indicative of predominant quenching/star-bursts.   

There is currently rich debate as to the shape of the $\sigma_{\mathrm{sSFR}}$-M$_{*}$ relation. At high redshifts ($z>1$) and for predominantly UV-derived SFRs, authors have generally found a relatively constant dispersion of $\sim0.3$\,dex \citep{Elbaz07, Noeske07, Rodighiero10,Whitaker12, Schreiber15}. However, other authors have suggested that this dispersion may increase with decreasing stellar mass at log$_{10}$[M$_{*}$/M$_{\odot}$]$<$10 in high redshift samples and be driven by stochastic SFHs in low mass galaxies \citep{Santini17}.  For more nearby samples and higher stellar mass galaxies other studies have identified a dispersion which increases at log$_{10}$[M$_{*}$/M$_{\odot}$]$>$10 \citep[][]{Guo15} and attribute the large dispersion to the presence of bulges and bars (their sample is purely selected based on sSFR and thus contain bulge+disc systems). To overcome this, \cite{Willett15} explore the $\sigma_{\mathrm{sSFR}}$-M$_{*}$ relation in a morphologically selected sample of disc-like spirals from Galaxy Zoo \citep{Willett13}. They find a minimum vertex parabolic (`U'-shaped) dispersion which decreases with stellar mass from log$_{10}$[M$_{*}$/M$_{\odot}$]$\sim$8-10 and then increases at log$_{10}$[M$_{*}$/M$_{\odot}$]$\sim$10-11.5. This is loosely consistent with the \cite{Guo15} results at the high mass end, but finds the upturn in dispersion occurs at higher masses. In addition, there also appears to be some variation in the literature measurements of $\sigma_{\mathrm{sSFR}}$-M$_{*}$ depending on the SFR indicator used; with UV-, H$\alpha$- and SED-derived SFRs producing different dispersions. This is potentially due to the varying biases included in different SFR indicators and the physical timescales over which they probe; for example H$\alpha$-derived SFRs are much more sensitive to short time-scale fluctuations in SFH \citep[for a detailed discussion of SFR indicators, their biases and timescales see][and Katsianis et al. 2017 for a simulations perspective]{Kennicutt12, Davies15b, Davies16b}.        

Evidently the observational picture is far from clear, with different teams applying different selection methods and using different SFR indicators finding different results. However, hydrodynamical simulations can offer some further insights into the $\sigma_{\mathrm{sSFR}}$-M$_{*}$ relation and the physical processes driving its shape. \cite{Sparre15} use the Illustris simulation \citep{Vogelsberger14} to explore the evolution of the sSFR-M$_{*}$ relation for all Illustris galaxies and at $z\sim0$ find a relatively flat $\sigma_{\mathrm{sSFR}}$-M$_{*}$ at  9$<$log$_{10}$[M$_{*}$/M$_{\odot}$]$<$10.5 and increasing dispersion to higher masses. However, they do not make any selection to exclude passive systems, and hence this increased dispersion is likely due to the passive population becoming more prevalent at high stellar masses.

More recently Katsianis et al (in preparation) applied a similar approach to the EAGLE simulation \citep{Crain15, Schaye15, McAlpine16, Matthee18} and find a minimum vertex parabolic (`U'-shaped) $\sigma_{\mathrm{sSFR}}$-M$_{*}$ relation similar to that of \cite{Willett15} - however, also see \cite{Matthee18} who find a linearly decreasing $\sigma_{\mathrm{sSFR}}$-M$_{*}$ in EAGLE. Moreover, the work of Katsianis et al also allows an exploration of the physical mechanisms which are driving this dispersion. First, they rerun their analysis with no-AGN feedback and find that the dispersion is dramatically reduced at the high stellar mass end; suggesting it is AGN feedback which drives the high dispersion at log$_{10}$[M$_{*}$/M$_{\odot}$]$\sim$10-11.5 observed by \cite{Guo15} and \cite{Willett15}. Next they rerun their analysis with stellar feedback turned off and find a reduced scatter at the low stellar mass end; suggesting it is stellar-feedback/star-formation which drives the observed dispersion at these lower masses. 

Similarly, using the semi-analytic model Shark, \cite{Lagos18} showed that the adopted star formation law had a strong effect on the scatter of the SFS (through their effect on the timescales of atomic to molecular hydrogen and molecular-to-stars conversion). However, the choice of star formation law rarely affected the zero-point of the main sequence. These simulation results suggest that the scatter of the SFS is rich in information about the physics of galaxy formation and provide us with both a prediction for the shape of the $\sigma_{\mathrm{sSFR}}$-M$_{*}$ relation and the physical mechanisms driving it. We must now aim to test this prediction via our observational samples.                            

In this series of papers we produce a detailed analysis of the sSFR-M$_{*}$ plane and the factors driving its formation. In this first paper we explore the observed dispersion along the SFS and its variation with sample selection, SFR indicator and group/isolated environment using the Galaxy And Mass Assembly (GAMA) sample. We explore whether the variation in the $z\sim0$  $\sigma_{\mathrm{sSFR}}$-M$_{*}$ observed by previous authors is physical or an artefact of their chosen method. Following this we will determine how the sSFR-M$_{*}$ plane can be subdivided into different populations based on various morphological and structural tracers, how this varies with environment, and how these populations are indicative of different evolutionary pathways through the sSFR-M$_{*}$ plane. Finally, we will use our new SED fitting code \textsc{prospect} (Robotham et al. in preparation) to determine SFHs and AGN fractions for galaxies across the sSFR-M$_{*}$ plane and explore the physical mechanisms which have shaped the $z\sim0$ SFR-M$_{*}$ relation using all observations presented in the series.       

This paper is organised as follows. In Section \ref{sec:data} we discus the observational samples used in this work and briefly describe our choice of SFR indicators. In Section \ref{sec:MS} we detail different methods for selecting the SFS based on SFR, colour and morphology/structure. In Section \ref{sec:massDep} we present the resultant $\sigma_{\mathrm{sSFR}}$-M$_{*}$ relation derived from each sample and explore its variation with SFR indicator and environment. We also compare our results to the Shark and EAGLE simulations. Finally, in Section \ref{sec:conclusions} we present our conclusions. Throughout this paper we use a standard $\Lambda$CDM cosmology with {H}$_{0}$\,=\,70\,kms$^{-1}$\,Mpc$^{-1}$, $\Omega_{\Lambda}$\,=\,0.7 and $\Omega_{M}$\,=\,0.3.

\begin{figure}
\begin{center}
\includegraphics[scale=0.45]{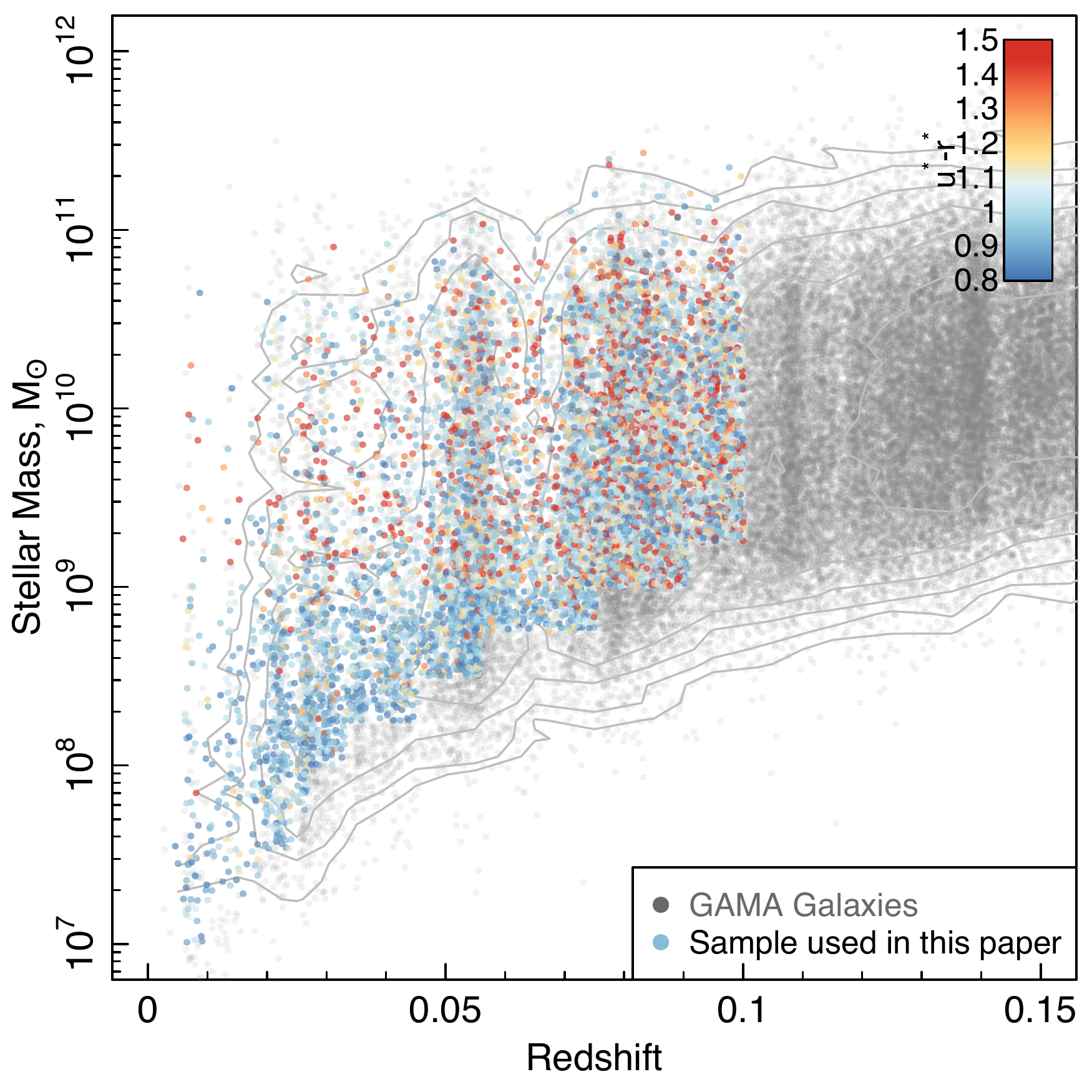}
\caption{The redshift-stellar mass distribution of galaxies used in this work in comparison to the full GAMA sample. We initially select isolated (non-group or pair) galaxies in volume-limited samples in log$_{10}$[M$_{*}$/M$_{\odot}$]=0.25 bins of stellar mass (see Section \ref{sec:GAMA} for details). Selected points are coloured by their rest-frame, extinction corrected $u-r$ colour. Contours display the density of GAMA points. }
\label{fig:Mz}
\end{center}
\end{figure}

\section{Data and Sample Selections}
\label{sec:data} 

\subsection{Galaxy And Mass Assembly Survey}
\label{sec:GAMA}

The GAMA survey second data release \citep[GAMA II][]{Liske15} covers 286\,deg$^{2}$ to a main survey limit of $r_{\mathrm{AB}}<19.8$\,mag in three equatorial (G09, G12 and G15) and two southern (G02 and G23 \- survey limit of $i_{\mathrm{AB}}<19.2$\,mag in G23) regions. The limiting magnitude of GAMA was initially designed to probe all aspects of cosmic structures on 1\,kpc to 1\,Mpc scales spanning all environments and out to a redshift limit of $z\sim0.4$. The spectroscopic survey was undertaken using the AAOmega fibre-fed spectrograph \citep[][]{Saunders04,Sharp06} in conjunction with the Two-degree Field \citep[2dF,][]{Lewis02} positioner on the Anglo-Australian Telescope and obtained redshifts for $\sim$240,000 targets covering $0<z\lesssim0.5$ with a median redshift of $z\sim0.25$, and highly uniform spatial completeness \citep[see][for summary of GAMA observations]{Baldry10,Robotham10,Driver11,Hopkins13}. 

Full details of the GAMA survey can be found in \citet{Driver11, Driver16},  \citet{Liske15} and \citet{Baldry18}. In this work we utilise the data obtained in the 3 equatorial regions, which we refer to here as GAMA II$_{Eq}$. Stellar masses for the GAMA II$_{Eq}$ sample, and those used in this work, are derived from the $ugriZYJHK$ photometry using a method similar to that outlined in \cite{Taylor11} - assuming a Chabrier IMF \citep{Chabrier03}. All photometry used in this work is measured using the Lambda Adaptive Multi-Band Deblending Algorithm for R (\textsc{lambdar}) and presented in \cite{Wright16}.

In this paper we also make use of the recent GalaxyZoo classifications that are based on the Kilo Degree Survey \citep[KiDS,][]{deJong13, deJong15, deJong17} imaging in the GAMA regions. For these classifications, 49,851 galaxies were selected from the GAMA equatorial fields with redshifts $z<0.15$. Within GalaxyZoo the GAMA sample received almost  2\,million classifications from over 20,000 unique users in 12 months. The GAMA-KiDS GalaxyZoo classifications use the standard decision tree implemented for current GalaxyZoo projects.  A full description of the GAMA-KiDS GalaxyZoo effort can be found in Kelvin et al. (in preparation).

In this work we further restrict our sample to galaxies within rolling volume-limited samples and initially galaxies that are not in either a group or pair in the GAMA group catalogues of \cite{Robotham11}; $i.e.$ isolated centrals. To define our volume-limited samples we follow a similar approach to \cite{Lange16} and split the full GAMA catalogue into $\Delta$log$_{10}$[M$_{*}$/M$_{\odot}$]=0.25 bins of stellar mass. For each bin we calculate the redshift where 97.7\% of the sample has a maximum observable redshift (V$_{\mathrm{max}}$) greater than the median V$_{\mathrm{max}}$ of the bin. We then exclude all galaxies above this redshift, within the particular stellar mass bin. Finally we put an upper redshift limit of $z<0.1$. This process allows us to extend to lower stellar masses in our very local sample.         

The resultant redshift-stellar mass selection of our sample in comparison to the full GAMA sample is shown in Figure \ref{fig:Mz}. The $z<0.1$ restriction is to largely remove any evolution in the SFS across our sample redshift range. In addition, a number of GAMA data products which are required for our analysis are only available for $z<0.1$ GAMA galaxies (see Section \ref{sec:MS}). The isolated centrals restriction allows us to remove any additional environmental quenching effects which may induce additional scatter in the SFS which is not driven by feedback \cite[however, $c.f.$][find that groups can affect galaxies SF properties out to large radii]{Barsanti18}. Note that we do not include group centrals in our sample as there is currently some debate as to whether or not these galaxies undergo environmental quenching \cite[$e.g.$ see][]{Wang18a}. However, we do also preform our analysis including group centrals and find that it does not significantly affect our results. The effect of group environment on $\sigma_{\mathrm{sSFR}}$ is then explored further in Section \ref{sec:group} and in the following papers in this series. In total there are  9005 galaxies in our starting sample.

\subsubsection{GAMA SFR indicators}
\label{sec:SFRindicators}

The GAMA SFR indicators used in this work are described at length in \cite{Davies16b}. Briefly, we use: i) \textsc{magphys}-derived SFRs outlined in \cite{Driver18} and based on the energy balance SED fitting code \textsc{magphys} \citep{daCunha08}, ii) combined Ultraviolet and Total Infrared (UV+TIR) SFRs derived from the \citet{Brown14} galaxy spectra, iii) H$\alpha$-derived SFRs using GAMA spectra discussed in \citet{Liske15} and the process outlined in \cite{Gunawardhana11,Gunawardhana15} and \cite{Hopkins13}, and using the line measurements of \cite{Gordon17}, iv) \textit{Wide-field Infrared Survey Explorer} ($WISE$) W3-band SFRs derived using the prescription outlined in \cite{Cluver17}, and v) extinction-corrected $u$-band SFRs derived using the GAMA rest-frame $u$-band luminosity and $u-g$ colours from \cite{Davies16b}. All SFRs are scaled to a Chabrier IMF and for further details of these SFR indicators and their derivation see \cite{Davies16b}. 

We repeat all of the analysis in this paper for each of these SFR indicators, but for clarity we initially only show results for \textsc{magphys}-derived SFRs. Similar figures for all of our indicators are presented in the Appendix and discussed in Section \ref{sec:massDep}. We note here that different SFR indicators can be appropriate for different science cases, and contain different biases/assumptions, as such we consider a variety here.

\section{Isolating the target population}
\label{sec:MS}

In order to explore the stellar mass dependance on $\sigma_{\mathrm{sSFR}}$, and in a similar manner to previous authors, we must first isolate our population of interest. The method by which this is undertaken is largely dependent on the specific scientific question being addressed \citep[$i.e.$ see][]{Renzini15}. For example, if we wished to study the dispersion within the SFS to explore self-regulated growth via star-formation, we may select sources based solely on SFR to exclude passive systems. Conversely, if we wish to investigate the SFH of star-forming discs we may wish to isolate morphologically-selected disc like systems, irrespective of their position in the sSFR-M$_{*}$ plane. However, care must then be taken when drawing inferences regarding the $\sigma_{\mathrm{sSFR}}$-M$_{*}$ relation when applying different selection methods, as these can significantly affect the observed distribution. For example, is the parabolic distribution observed by previous authors largely driven by sample selection and not pure physical processes? 

Here we aim to explore the impact of sample selection on the $\sigma_{\mathrm{sSFR}}$-M$_{*}$ relation and thus provide a robust description of the intrinsic shape of the dispersion. In the following subsections we explore a number of different selection criteria for identifying sources which may be used to parameterise $\sigma_{\mathrm{sSFR}}$-M$_{*}$. In subsequent sections, the impact of each of these selections on the shape of the derived $\sigma_{\mathrm{sSFR}}$ relation will be explored. The populations selected by each of these selections is displayed in Figure \ref{fig:selections} for \textsc{magphys}-derived sSFRs.  In Section \ref{sec:mix} we will also explore using a non-physically motivated mixture-modelling method to determine the $\sigma_{\mathrm{sSFR}}$-M$_{*}$, but separate it from the physically motivated selection applied here.   

\subsection{No Selection}
\label{sec:none}

Firstly we explore the $\sigma_{\mathrm{sSFR}}$-M$_{*}$ relation with no selection applied to the population (other than those previously described). This sample contains all 9005 sources with both star-forming and passive systems, and morphological pure discs, ellipticals and two component disc+bulge systems. The $\sigma_{\mathrm{sSFR}}$-M$_{*}$ relation derived from this distribution is most directly comparable to the previous simulation results from Illustris and EAGLE which do not apply any selection criteria; largely because all other selections are non-trivial to compute using the simulation data. In addition, this selection is informative in its own right as it essentially describes global SFH of all galaxies at a given stellar mass and can be used to identify star-burst/quenched populations irrespective of their non-star-forming properties.  Panel A of Figure \ref{fig:selections} displays our sample with no further selection for \textsc{magphys}-derived SFRs (and in the appendix for all other SFR indicators).

\subsection{$u-r$ Colour Selection}
\label{sec:col}

One potential method for identifying late-type (star-forming) galaxies is through rest-frame, extinction-corrected broad-band optical colours. Galaxy colours have a long history in selecting star-forming, passive and green-valley systems \cite[for some examples see][]{Salim05,Schawinski14,Taylor15}. Old stellar populations show a red colour with substantially more flux at longer wavelengths. However, with star-formation activity high-mass, blue stars increasingly contribute to the galaxy's spectral shape, flattening the SED and producing bluer colours. As such, optical colour provides a measure of recent star-formation history and can be used to isolate star-forming systems. Here we use $u-r$ colour to isolate late-type, star-forming galaxies as the $u$-band is strongly correlated with recent star-formation \citep[see][]{Davies16b} and the $r$-band is representative of the underlying older stellar population. 

We use the GAMA extinction-corrected rest-frame $u^*-r^*$ colours taken from \cite{Taylor11} and separate the populations following the approach outlined in \cite{Bremer18}, where we apply a mass-dependent colour selection for star-forming galaxies of:

\begin{equation}
u^*-r^*<0.15 \times \mathrm{log}_{10}[\mathrm{M}_{*}/\mathrm{M}_{\odot}]+0.05
\end{equation}

\noindent This relation sits between the blue and red galaxy selection lines of \cite{Bremer18}. This forms our $u-r$ colour selected sample and is displayed in panel B of Figure \ref{fig:selections}. 8338 sources are selected star-forming using this criteria, while 667 are classed as passive. We note that a $u-r$ colour selection is not as extreme as a pure sSFR cut (described in the next section) as the $u-r$ colour essentially probes the contribution of young stellar populations over a relatively large evolutionary baseline, while sSFR measures the recently formed stellar population on a short timescale. In addition, the $u-r$  colour selection is sensitive to dust obscuration. While we use extinction-corrected colours, these my suffer from large uncertainties in extremely dusty galaxies.

\subsection{sSFR Selection}

\begin{figure*}
\begin{center}
\includegraphics[scale=0.47]{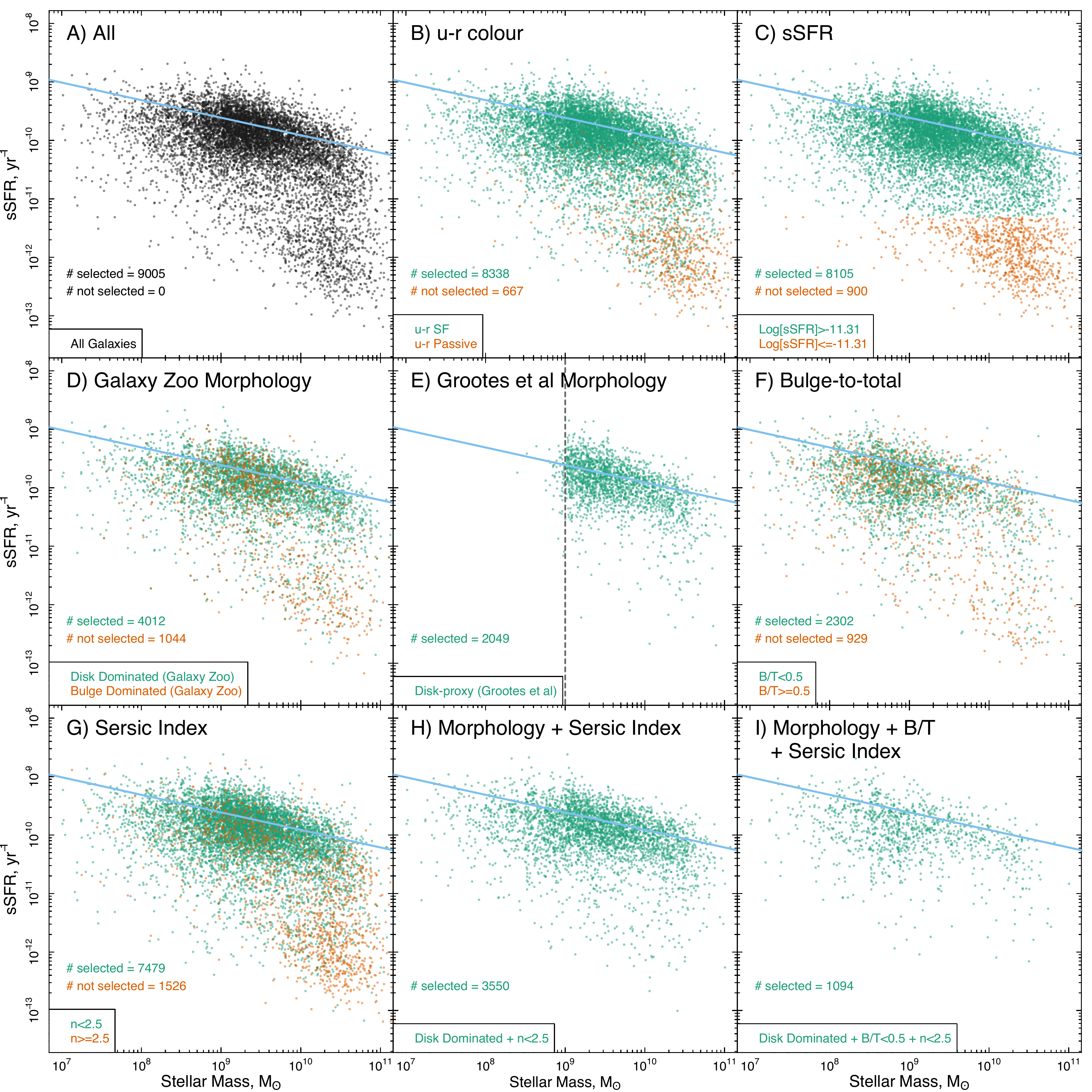}
\caption{The sSFR$_{\mathrm{MAGPHYS}}$-M$_{*}$ relation colour-coded by various selection methods for identifying the star-forming `sequence' (SFS). The linear fit (light blue line) is taken from \citet{Davies16b}. A: the full distribution of $z<0.1$ GAMA galaxies which are not in a group or pair ($i.e.$ isolated down to GAMA limit), B: selection based on $u-r$ colour, C: selection based on sSFR, D: selection based on visual morphological classification from Galaxy Zoo, E: selection based on the morphological proxy analysis of \citet{Grootes14} - vertical dashed line shows stellar mass limit of this analysis, F: selection based on r-band bulge-to-total light ratio (B/T) - note this sample only extends to $z<0.06$, G: selection based on S\'{e}rsic index, H: selection based on visual morphology and S\'{e}rsic index selection combined, I) selection based on visual morphology, S\'{e}rsic index and B/T selections combined. The number of galaxies in each selection are displayed in the bottom left corner of each panel. The total number of selected and non-selected galaxies displays the number of sources for which we have the required measurements for the selection. For panels which do not have a value for `non-selected', either the catalogue used only contains the selected sample (panel E), or the selection if knowingly incomplete and the `non-selected' number is not informative (panels H and I).  }
\label{fig:selections}
\end{center}
\end{figure*}

It is also possible to isolate the SFS using star-formation alone. To do this one must somewhat arbitrarily choose at which SFR to divide the two populations, which can have a strong effect on the measured dispersion ($i.e.$ too low a cut and you include the passive population, too high a cut and you artificially reduce $\sigma_{\mathrm{sSFR}}$). In addition, applying a single sSFR cut to isolate the star-forming population is only appropriate if the slope of the sSFR-M$_{*}$ relation is unity, this is not the case (see panel C of Figure \ref{fig:selections}). However, given that this approach is applied by many previous studies \citep[$i.e.$][]{Guo15}, we replicate a similar selection here. With the caveats of potential errors due to choice of star forming-passive cut in mind, here we isolate the SFS by selecting galaxies for which log$_{10}$[sSFR, yr$^{-1}$]$>$SFS$_{10}$-ScSFR\,dex. Where SFS is the fit to sSFR-M$_{*}$ taken from \cite{Davies16b} - shown as the blue line in Figure \ref{fig:selections}, SFS$_{10}$ is the normalisation of the fit at log$_{10}$[M$_{*}$/M$_{\odot}$]$=$10 and ScSFR=1.4,0.8,0.8,0.9, and 0.6 for \textsc{magphys}, UV+TIR, H$\alpha$, W3 and u-band SFRs respectively (to account for varying scatter/normalisation on the relations). The impact of this selection for \textsc{magphys} is displayed in panel C of Figure \ref{fig:selections}.  For \textsc{magphys} SFRs,  8105 sources are selected star-forming, while 900 are classed as passive.          

\subsection{Morphological/Structural Selections}
\label{sec:morp}

In order to produce samples which may be used to explore the recent SFH of spiral/disc-like systems, we also use a number of different morphological selections. These selections are aimed at identifying disc-dominated systems irrespective of their position relative to the locus of the SFS, and in practice may be used to explore the evolution of all disc-like galaxies in the SFR-M$_{*}$ plane.

\subsubsection{Galaxy Zoo}

Our initial disc-like selection is based on the GAMA-KiDS Galaxy Zoo classifications outlined in Section \ref{sec:GAMA}. For disc-dominated systems we select galaxies which are not edge-on (so that visual classifications are not confused) and have no bulge, or are classified as a spiral:

\begin{equation}
(p_{\mathrm{no-bulge}}>0.619 \\ \mathrm{\&} \\ p_{\mathrm{not-edge-on}}>0.715) \\ \mathrm{|} \\ p_{\mathrm{spiral}}>0.619
\end{equation}

\noindent where $p$ is the debiased vote fraction. For comparison in panel D of Figure \ref{fig:selections} we also show bulge-dominated systems selected as:

\begin{equation}
p_{\mathrm{bulge-dominant}}>0.619 \\ \mathrm{\&} \\ p_{\mathrm{not-edge-on}}>0.715. 
\end{equation}

\noindent  This selection is based around those used by the Galaxy Zoo team in \cite{Willett15} for spiral galaxies. Note that we do explore various other $p$ values for each of the selections above and find that within sensible ranges it does not have a strong impact on the shape of the derived $\sigma_{\mathrm{sSFR}}$-M$_{*}$ relation. Using these selections,  4012 galaxies are classed as disc-dominated and 1044 as bulge-dominated. The remaining 3949 are two-component or ambiguous, and are excluded.

\subsubsection{Grootes et al disc Proxy}

\cite{Grootes14} used a non-parametric cell-based method to identify a highly complete and pure sample of spiral galaxies at log$_{10}$[M$_{*}$/M$_{\odot}$]$>$9.0. In this approach colour, S\'{e}rsic index, stellar mass surface density and effective radius from SDSS imaging were used as a proxy for morphological classifications. These were then compared to previous SDSS-based Galaxy Zoo classifications to highlight the robustness of their method. Here we use the Grootes et al sample as spiral systems which reside on the SFS. However, we highlight that these morphological proxies only extend down to log$_{10}$[M$_{*}$/M$_{\odot}$]$\sim$9.0 and therefore do not cover our full sample (see panel E of Figure \ref{fig:selections}). There are 2049 source in the Grootes et al catalogue which meet our initial selections.

\subsubsection{Bulge-to-Total Mass}

In addition to morphological selections, we can also select samples based on a galaxy's structural components. These components represent a largely non-subjective measure of structural evolution from pure disc-like systems, to two component bulge+disc, and pure spheroid ellipticals. A simple metric for these classifications is the galaxy's bulge-to-total flux ratio (B/T); the ratio of light arising from the bulge compared to total bulge+disc where B/T=1 is a pure elliptical and B/T=0 is a pure disc. Using the 2-component S\'{e}rsic fits to GAMA galaxies based on SDSS $r$-band imaging \citep{Lange16} we select a population with B/T$<$0.5 as disc-dominated systems (see panel F of Figure \ref{fig:selections}). The analysis of \cite{Lange16} only provides 2-component S\'{e}rsic fits at $z<0.06$ leaving a starting sample of 3231 galaxies, 2302 of which have B/T$<$0.5 and 929 B/T$>=$0.5. Thus, this selection is limited to local galaxies. Note that we repeat our analysis with a more extreme B/T$<$0.3 selection to exclude equally-weighted bulge-disc systems, but find similar results.

\subsubsection{S\'{e}rsic Index}

A cruder metric for galaxy structure is the single component S\'{e}rsic index ($n$).  Briefly, the S\'{e}rsic index is a measure of the shape of a galaxy's light profile, with $n$=1 following an exponential profile, which is found to fit galaxy discs, and $n$=4 a de Vaucouleurs profile commonly associated with spheroidal components, such as bulges and elliptical galaxies. This is simply an empirical formalism which is found to fit the two-dimensional light profile of a galaxy but does not fully represent the three-dimensional distribution of stars. \cite{Kelvin12} produced single-S\'{e}rsic fits for GAMA galaxies using the SDSS and UKIDSS LAS imaging bands outlined in \cite{Hill11}. Here we use the SDSS $r$-band S\'{e}rsic index to isolate a disc-like sample with an $n<2.5$ selection. This selection is found to separate disc-dominated and spheroid-dominated systems in \cite{Lange15}. Panel G of Figure \ref{fig:selections} displays our S\'{e}rsic index selected sample isolating the SFS and removing galaxies in the passive cloud. 7478 sources are selected as low S\'{e}rsic index using this criteria, while 1526 are classed as having high S\'{e}rsic index.

\subsubsection{Combined Morphological Selections}

In addition to our individual morphological selections, we also produce samples using combined Galaxy Zoo morphology and single S\'{e}rsic Index (Panel H of Figure \ref{fig:selections}, selecting 3550 sources) and Galaxy Zoo morphology, single S\'{e}rsic Index and B/T (Panel I of Figure \ref{fig:selections}, selecting 1094). These samples are less complete, but form a more robust selection of disc-like systems than the individual morphological selections as they must meet 2 or more of our selection criteria. We do not show the `not selected' galaxies here, as the selection is incomplete and therefore, the `not selected' population is not informative.

\section{A mass dependent $\sigma_{\mathrm{sSFR}}$?}
\label{sec:massDep}

\begin{figure*}
\begin{center}
\includegraphics[scale=0.52]{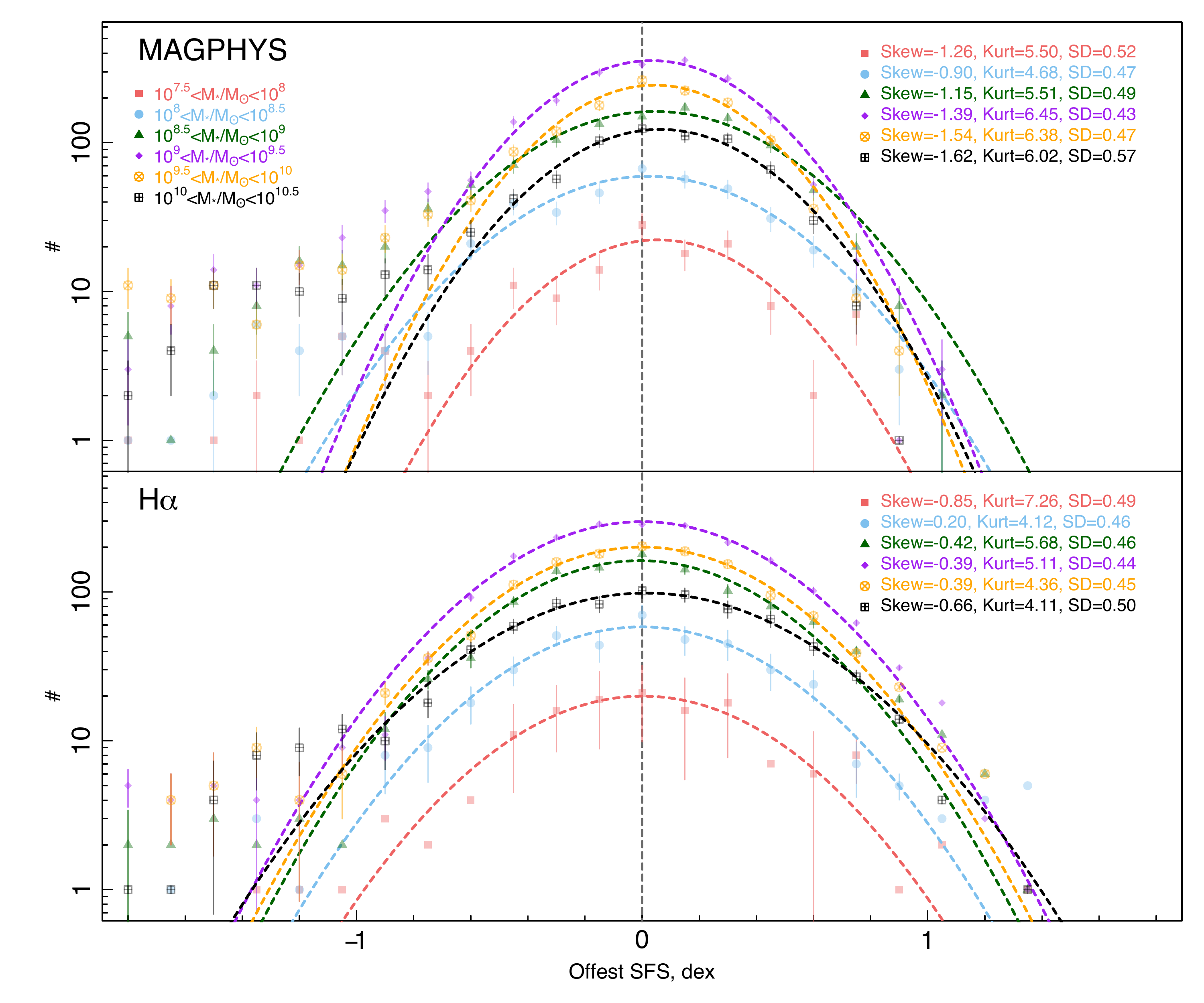}
\caption{The distribution of points away from the SFS isolated using the combined morphology+S\'{e}rsic index selection from Section \ref{sec:morp}. We display the distribution in stellar mass bins for both \textsc{magphys} and H$\alpha$ derived SFRs. Points are fit with a log-normal distribution (dashed lines) to highlight any deviation from this symmetric shape. In both SFRs and at all masses the distributions display a negative skew value, highlighting that the distributions have asymmetric tails extending to low stellar masses. For both SFRs in all but the lowest stellar mass bin the skew value becomes more negative to higher stellar masses suggesting that the distribution becomes more asymmetric as a function of mass.}
\label{fig:assym}
\end{center}
\end{figure*}

Using each of the sample selection methods described above, we explore $\sigma_{\mathrm{sSFR}}$ as a function of stellar mass. Here we define $\sigma_{\mathrm{sSFR}}$ as the standard deviation of log$_{10}$[sSFR] for the selected population at a given stellar mass. Note that following this we also calculate the dispersion based on the interquartile range of log$_{10}$[sSFR], to remove any dependencies on assumption of a Gaussian-like distribution. 

We take all of our selected galaxies (black points in Figure \ref{fig:selections} A, and green points in Figure \ref{fig:selections} B-I) and split our samples into log$_{10}$[M$_{*}$/M$_{\odot}$]=0.5 bins of stellar mass. Figure \ref{fig:assym} displays an example of the distribution of galaxies away from the median in each bin with Poisson errors, for both \textsc{magphys} and H$\alpha$ SFRs. For this figure we only show sources identified as disc-like from our combined Galaxy Zoo morphology and S\'{e}rsic index selection ($i.e.$ the population in Panel H of Figure \ref{fig:selections}). This shows the spread of sSFRs away from the locus of the SFS at a given stellar mass.  

\begin{figure*}
\begin{center}
\includegraphics[scale=0.47]{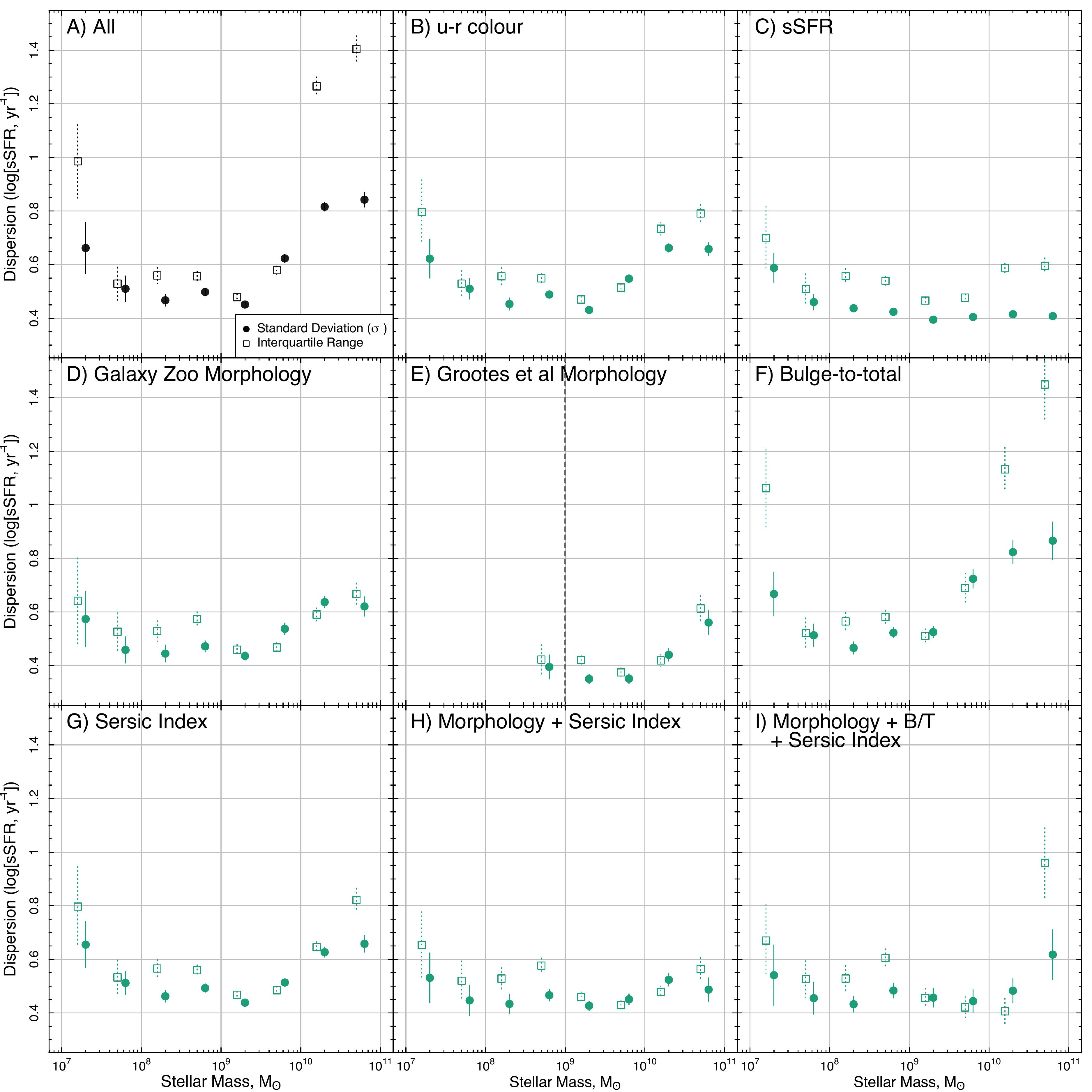}
\caption{The resultant dispersion along the sSFR-M$_{*}$ relation measured using both the standard deviation ($\sigma_{\mathrm{sSFR}}$-M$_{*}$, filled circles) and interquartile range (open squares). Each panel shows a different method for isolating the SFS described in Section \ref{sec:MS}.  This Figure follows the same layout as Figure \ref{fig:selections}, but only using SFS-selected populations (green points when both green and red are present in Figure \ref{fig:selections}). While the normalisation and shape changes, the majority distributions show a `U'-shaped  dispersion with minimum at $\sim10^9.25$\,M$_{\odot}$ and increased scatter at low and high masses for both $\sigma_{\mathrm{sSFR}}$ and interquartile range.    }
\label{fig:scatter}
\end{center}
\end{figure*}

  \begin{figure}
\includegraphics[scale=0.5]{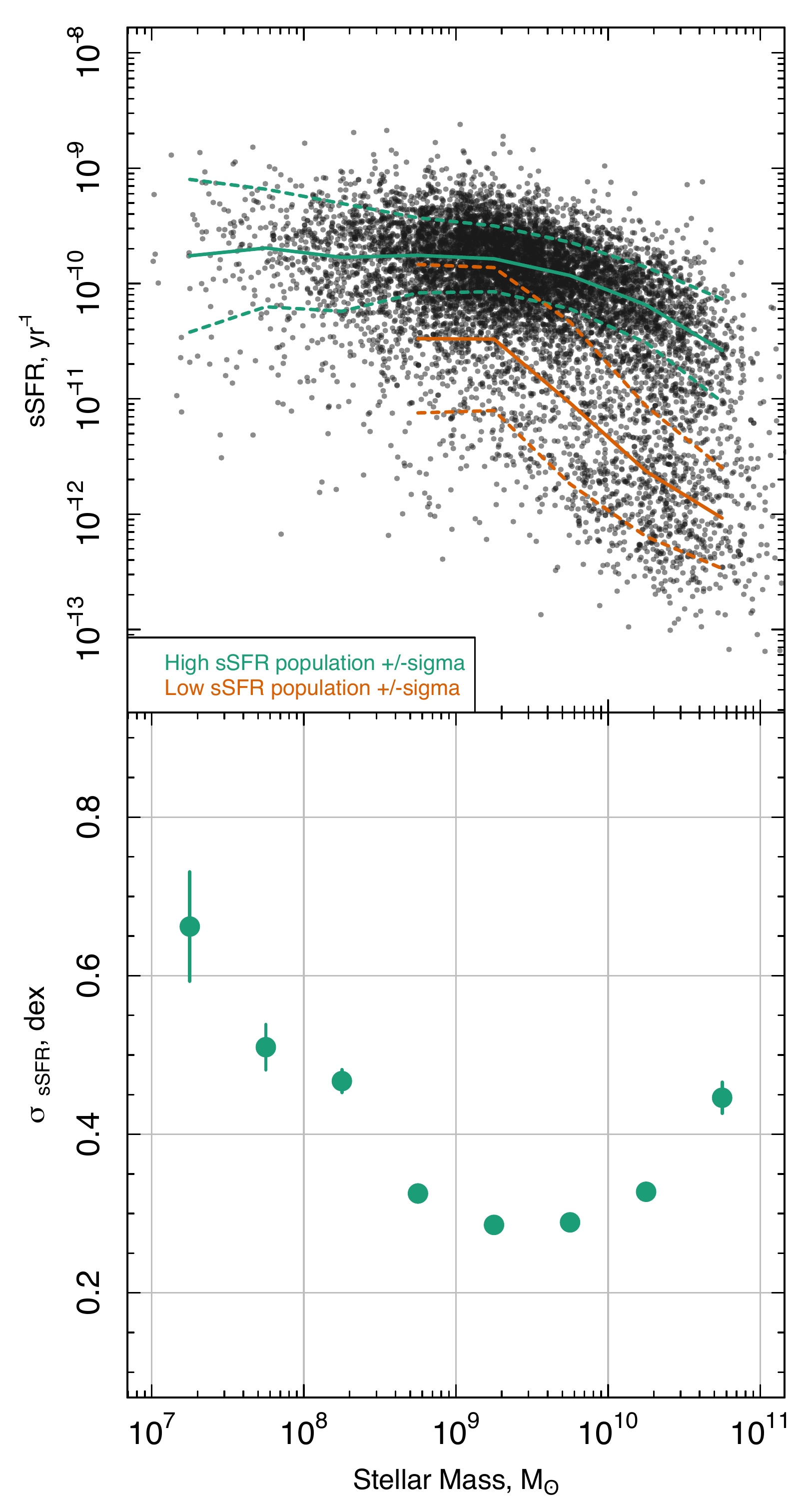}
\caption{Mixture modelling of the star-forming and passive populations assuming log-normal distributions in sSFR. The top panel displays the SFR-M$_{*}$ plane for \textsc{magphys}-derived SFRs. The green and red lines display the mean (solid lines) and 1$\sigma$ width (dashed lines) of the mixture models as a function of stellar mass. The bottom panel show the $\sigma_{\mathrm{sSFR}}$-M$_{*}$ relation for the high sSFR population.}
\label{fig:mix}
\end{figure}

For illustrative purposes only, in Figure \ref{fig:assym} we fit a log-normal in each mass bin (dashed lines) to highlight any deviation in the populations from a symmetric log-normal distribution. Unsurprisingly at all masses we see an asymmetric tail extending to low sSFRs indicative of a green-valley/passive population \citep[also see][]{Oemler17}. We also potentially see an additional star-burst population in the H$\alpha$ distribution at high SFRs (particularly seen in the positive kurtosis of the 8$<$log$_{10}$[M$_{*}$/M$_{\odot}$]$<$9 bin). This is interesting as it may provide evidence of short duration star-burst activity; as H$\alpha$ probes star-formation integrated over shorter time-scales than \textsc{magphys} \citep[see][]{Davies15b}. This will be explored further in later papers of this series.     

To determine how these distributions vary as a function of stellar mass, we calculate the skewness, kurtosis and standard deviation of log$_{10}$[sSFRs] in each mass bin; given in the legend of Figure \ref{fig:assym}. Interestingly, with the exception of the 8.0$<$log$_{10}$[M$_{*}$/M$_{\odot}$]$<$8.5 bin in H$\alpha$, all distributions show a negative skew (from the green valley/passive population) but with skewness increasing for higher stellar masses for bins at log$_{10}$[M$_{*}$/M$_{\odot}$]$>$8.0. Tentatively, this may indicate that any physical processes which drive the shape of this distribution are more likely to produce more symmetrical scatter at the low mass end (potentially stochastic star-formation bursts leading to stellar feedback), and asymmetrical scatter at the high mass end (potentially AGN feedback). However, care must be taken as we may simply be missing a significant fraction of the quenched population at low stellar masses, and artificially removing the asymmetric tail due to selection affects imposed by GAMA. In order to go further we require much deeper spectroscopic surveys of the local Universe, such as the Wide Area VISTA Extragalactic Survey \citep[WAVES][]{Driver16b}, which will allow a detailed exploration of the distribution of star-formation in very low mass galaxies.

  \begin{figure*}
\begin{center}
\includegraphics[scale=0.47]{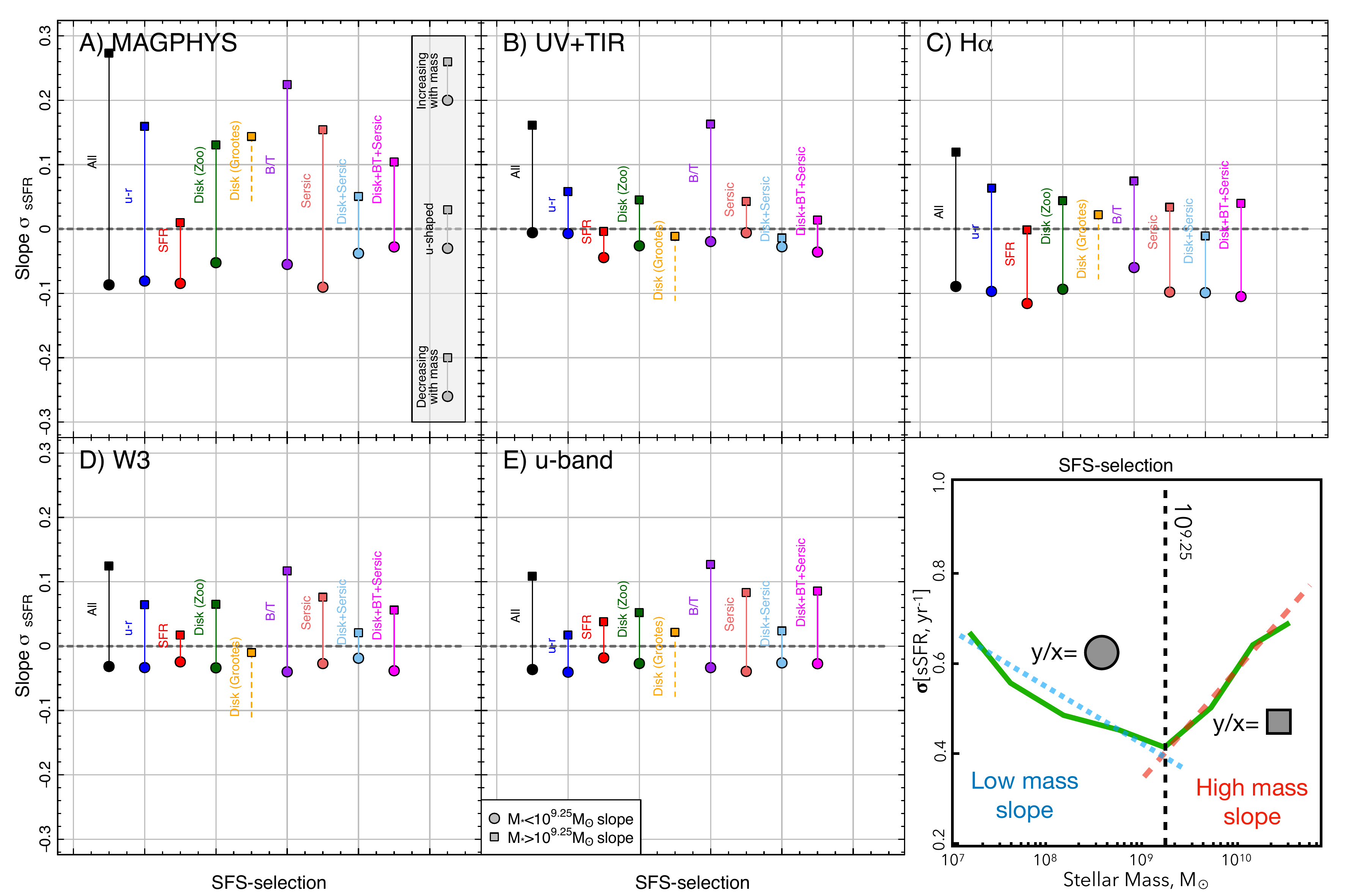}
\caption{The high-mass and low-mass slope of the $\sigma_{\mathrm{sSFR}}$-M$_{*}$ relation for different SFR indicators (described in Section \ref{sec:SFRindicators}) and using each method for isolating the SFS (described in Section \ref{sec:MS}).  We separate the $\sigma_{\mathrm{sSFR}}$-M$_{*}$ distributions in Figure \ref{fig:scatAll} and take the high-mass end as log$_{10}$[M$_{*}$/M$_{\odot}$]$>$9.25 and low-mass end as log$_{10}$[M$_{*}$/M$_{\odot}$]$<$9.25. We then linearly fit the slope at high- (squares) and low-masses (circles) - see bottom right panel. Each other panel displays a different SFR indicator, and each pair of points/line displays a different method for isolating the SFS. In this figure when the pair of points span the dashed zero line, the distribution $\sigma_{\mathrm{sSFR}}$-M$_{*}$ distribution is parabolic with a minimum vertex, when they both sit below the dashed zero line $\sigma_{\mathrm{sSFR}}$ decreases with mass, and when they both sit above the dashed zero line $\sigma_{\mathrm{sSFR}}$ increases with mass. The size of the distance between the two points is indicative of how curved the distribution is about log$_{10}$[M$_{*}$/M$_{\odot}$]$=$9.25. Note that for the Grootes et al morphological proxy, we only have the high mass slope. We find that in almost all cases the $\sigma_{\mathrm{sSFR}}$-M$_{*}$ relation is parabolic  (the pairs of points span the dashed zero line). We also note that if we use either log$_{10}$[M$_{*}$/M$_{\odot}$]$=$8.75 or log$_{10}$[M$_{*}$/M$_{\odot}$]$=$9.75 as our minimum points it does not significantly alter these results. }
\label{fig:slopes}
\end{center}
\end{figure*}

Taking this further we then use the standard deviation measurements described above to derive the $\sigma_{\mathrm{sSFR}}$-M$_{*}$ and measure the interquartile range in the same stellar mass bins; for all selections described in Section \ref{sec:MS}. 

Figure \ref{fig:scatter} displays $\sigma_{\mathrm{sSFR}}$ (filled circles) and the interquartile range (open squares) in log$_{10}$[M$_{*}$/M$_{\odot}$]=0.5 bins of stellar mass following the same panel layout as Figure \ref{fig:selections}.  Here we only show figures for \textsc{magphys} SFRs, but similar figures for other SFR indicators are given in the Appendix. Statistical errors on $\sigma_{\mathrm{sSFR}}$ are calculated as:

\begin{equation}
\label{eq:errors}
\mathrm{Err}_{\sigma_{\mathrm{sSFR}_i}}\sim\frac{\sqrt{2\sigma_{\mathrm{sSFR}_i}^4 (N_i-1)^{-1}}}{2\sigma_{\mathrm{sSFR}_i}}
\end{equation}

\noindent where $i$ is the index of the stellar mass bin and $N$ is the number of galaxies in that bin \citep{Rao73}. We then combine this in quadrature with the error calculated from 100 bootstrap resamples of the population within the errors of both stellar mass and SFR. For errors on the interquartile range we only apply the bootstrap resampling errors. This bootstrap resampling in intended to take into account the varying measurement error in each of our SFR indicators as a function of stellar mass.    

Firstly, we find that the majority of the panels show close to `U'-shaped distribution with minima at $\sigma_{\mathrm{sSFR}}\sim0.35-0.5$\,dex and log$_{10}$[M$_{*}$/M$_{\odot}$]=8.5-9.5 (in the following section we fit these distributions with a $2^{nd}$ order polynomial to find the minimum points). We observe a steep increase in both $\sigma_{\mathrm{sSFR}}$ and interquartile range to higher stellar masses than this minima. The increase to lower stellar masses is much shallower but tentatively observable in both measures of dispersion.  These results are roughly consistent with observations of \cite{Guo15} and \cite{Willett15} modulo slight differences in normalisation ($\sim0.1$\,dex). We also find a consistency between the dispersion measured using both $\sigma_{\mathrm{sSFR}}$ and the interquartile range at most stellar masses, this indicates that the assumption of a log-normal distribution of sSFRs at a given stellar mass is appropriate. The places where this does not hold true are when stellar masses are either very low or very high (this is also displayed in the larger kurtosis values in Figure \ref{fig:assym}). In these regimes, the interquartile range may be a more accurate representation of the dispersion as it does not assume a log-normal distribution.        

We find that while sample selection does affect the normalisation of the dispersion at specific stellar masses, we do see consistency between sample selections in terms of the minimum point in both stellar mass and dispersion, and in the overall shape; with all samples displaying a dispersion which rises to low and high stellar masses.    

We also find that when selecting on sSFR (panel C) we observe a more linear distribution with decreasing $\sigma_{\mathrm{sSFR}}$ to higher stellar masses. This is as expected as a hard cut in sSFR simply removes the high dispersion population at high stellar masses (see panel C of Figure \ref{fig:selections}). In combination the panels in Figure \ref{fig:scatter} suggests that irrespective of the method for selecting a star-forming/disc-like galaxy population based on colour and morphology, in GAMA we observe a `U'-shaped $\sigma_{\mathrm{sSFR}}$-M$_{*}$ relation which is largely consistent with the previous observations of \cite{Guo15} and \cite{Willett15}.   

\subsubsection{Mixture Modelling Method}     
\label{sec:mix}

In addition to the physically-motivated sample selections described previously, it is also possible to determine the $\sigma_{\mathrm{sSFR}}$-M$_{*}$ relation based on mixture modelling of the star-forming and passive populations \citep[see][for a similar approach]{Taylor15}. Here we use the [R] \textsc{mixtools:normalmixem} function to perform a maximum log-likelihood mixture fit to the two populations in log$_{10}$[M$_{*}$/M$_{\odot}$]=0.5 bins of stellar mass at log$_{10}$[M$_{*}$/M$_{\odot}$]$>$8.5 (below this we assume a single high sSFR population). The top panel of Figure \ref{fig:mix} shows the sSFR-M$_{*}$ plane for \textsc{magphys}-derived SFRs. The green and red lines display the mean (solid lines) and 1$\sigma$ width (dashed lines) of the mixture models for the high sSFR and low sSFR population respectively.  

The bottom panel of Figure \ref{fig:mix} then shows $\sigma_{\mathrm{sSFR}}$-M$_{*}$ relation for the high sSFR population, where errors are calculated using Equation \ref{eq:errors} but where $N$ is the number of galaxies in a particular stellar mass bin multiplied by the mixing proportion in the high sSFR population. Once again we find a `U'-shaped distribution with minima at log$_{10}$[M$_{*}$/M$_{\odot}$]$\sim$9.25 and $\sigma_{\mathrm{sSFR}}$ increasing to low and high stellar masses. We do find that the normalisation of the $\sigma_{\mathrm{sSFR}}$-M$_{*}$ relation for our mixtures is slightly lower than for our previous selections (minima at $\sigma_{\mathrm{sSFR}}\sim0.3$\,dex), likely due to the fact that large dispersion points are included in the low sSFR mixture irrespective of their physical properties.

However, it is interesting to highlight that when using non-physically motivated separation between the star-forming and passive population we still observe a `U'-shaped $\sigma_{\mathrm{sSFR}}$-M$_{*}$. This result is therefore once again, unlikely to be driven by choice of sample selection.

\subsection{Variation with SFR indicator}     
\label{sec:var}

We also consider if the results described above hold true for different SFR indicators; as different observables can provide different measures of star-formation with varying degrees of scatter \citep[$e.g$ see][]{Davies16b}. In addition, other explorations into the dispersion along the SFS have used a number of different SFR indicators. As such, we wish to produce a comparison point to these studies from GAMA in the local Universe.    

Hence, we reproduce our analysis for all other SFR indicators discussed in Section \ref{sec:SFRindicators}. Figures identical to Figures \ref{fig:selections} and \ref{fig:scatter} are given in the appendix for UV+TIR, H$\alpha$, W3 and U-band SFRs. In the majority of cases these figures display a largely `U'-shaped $\sigma_{\mathrm{sSFR}}$-M$_{*}$ relation with minimum at $\sigma_{\mathrm{sSFR}}\sim0.35-0.5$\,dex and log$_{10}$[M$_{*}$/M$_{\odot}$]=9-10. They all display a steep increase of dispersion to high stellar masses, and a more tentative, shallow increase to lower stellar masses. 

The u-band SFRs (Figure \ref{fig:scatterU}) do have a much smaller dispersion across all stellar masses (minimum at $\sigma_{\mathrm{sSFR}}\sim0.2-0.3$\,dex) but retain a `U'-shape.  The changes in normalisation of $\sigma_{\mathrm{sSFR}}$ between SFR indicator are likely due to the robustness of each indicator in measuring SFRs and how correlated they are with stellar mass; with high scatter indicative of larger measurement error and/or unknown assumptions in converting from an observable to a true SFR\citep[$e.g$ see][]{Davies16b}. 

In Figure \ref{fig:scatAll}, we produce a comparison of all SFR indicators by scaling the distributions using the minimum dispersion in each indicator. This removes some of the dependancy on absolute normalisation between SFR indicators and allows a more direct comparison of their shape. In the majority of cases each indicator shows a `U-shape' with largely consistent increases in dispersion to low and high stellar masses, irrespective of SFR used. The most notable exception to this is the UV+TIR indicator. This is largely due to the fact the the minimum point in $\sigma_{\mathrm{sSFR}}$ occurs at higher stellar masses for this indicator (close to log$_{10}$[M$_{*}$/M$_{\odot}$]=10).

\begin{figure*}
\begin{center}
\includegraphics[scale=0.45]{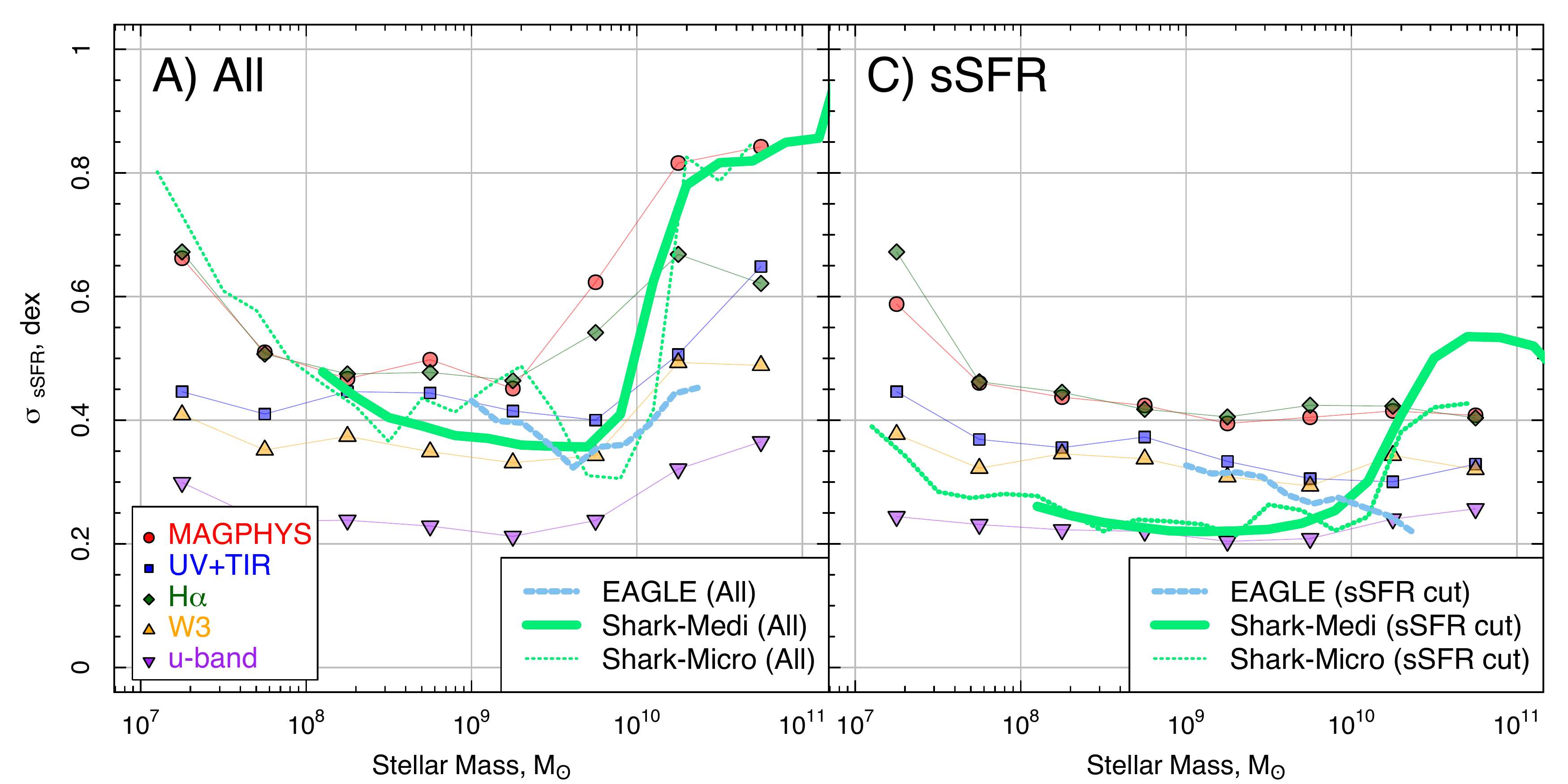}
\caption{$\sigma_{\mathrm{sSFR}}$-M$_{*}$ relation from the Shark semi-analytic model (solid green) compared to our observations for different SFR indicators (coloured points) and the EAGLE simulation results (dashed blue lines). The left panel displays the relation for all galaxies, while the right shows only galaxies with log$_{10}$[sSFR, yr$^{-1}$]$>$-11. Both simulations are only shown above their mass resolution limits. Shark displays a remarkably similar distribution to our data in the left panel, and both Shark and EAGLE show a similar relation but offset slightly in normalisation in the right panel at log$_{10}$[M$_{*}$/M$_{\odot}$]<10. At the highest stellar masses we see an upturn for Shark which is not found in EAGLE or the observations; due to that fact that in Shark, star formation in very massive galaxies declines slowly and galaxies stay above the sSFR cut. }
\label{fig:shark}
\end{center}
\end{figure*}

\begin{figure*}
\begin{center}
\includegraphics[scale=0.35]{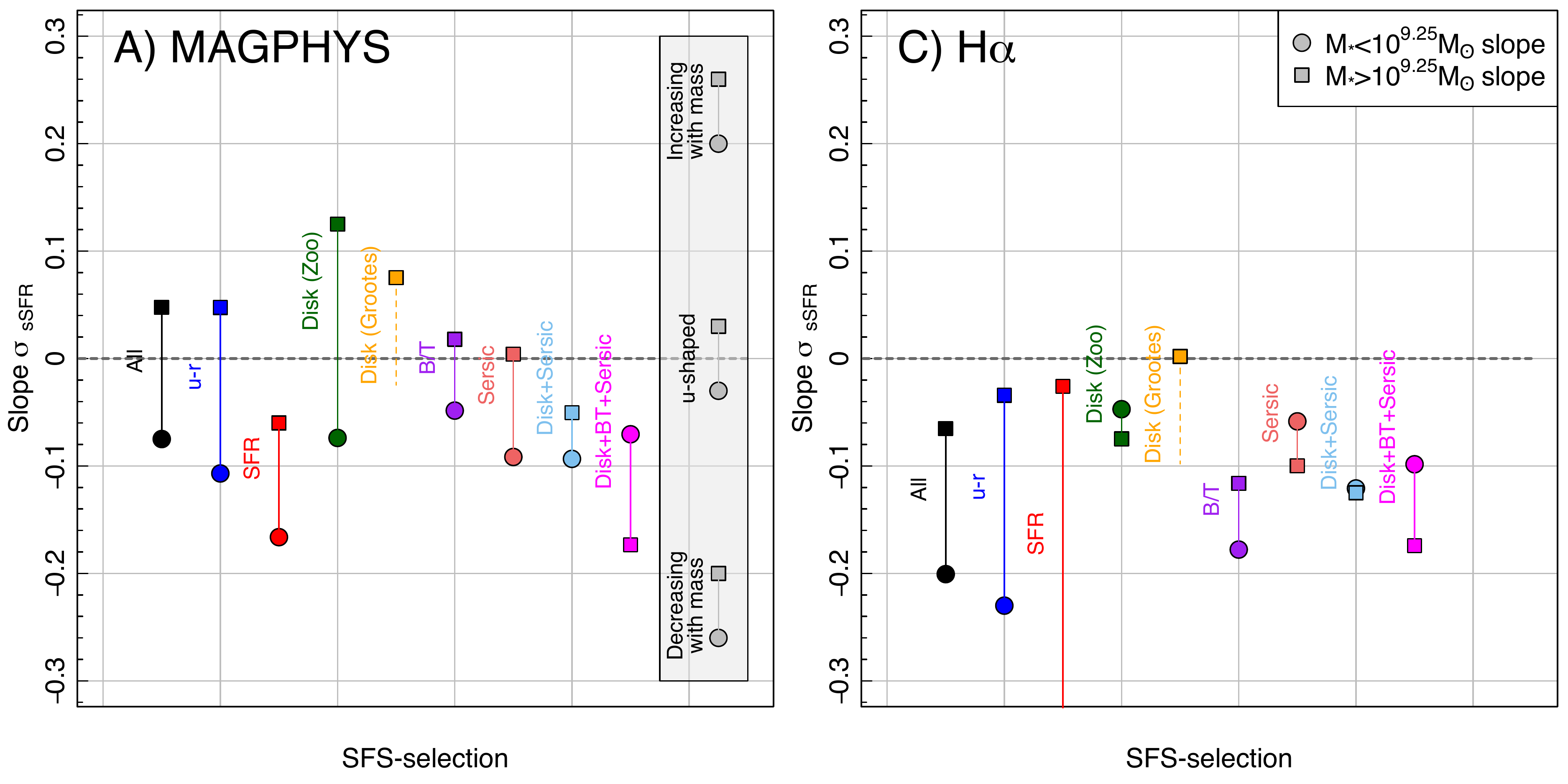}
\vspace{-5mm}

\includegraphics[scale=0.35]{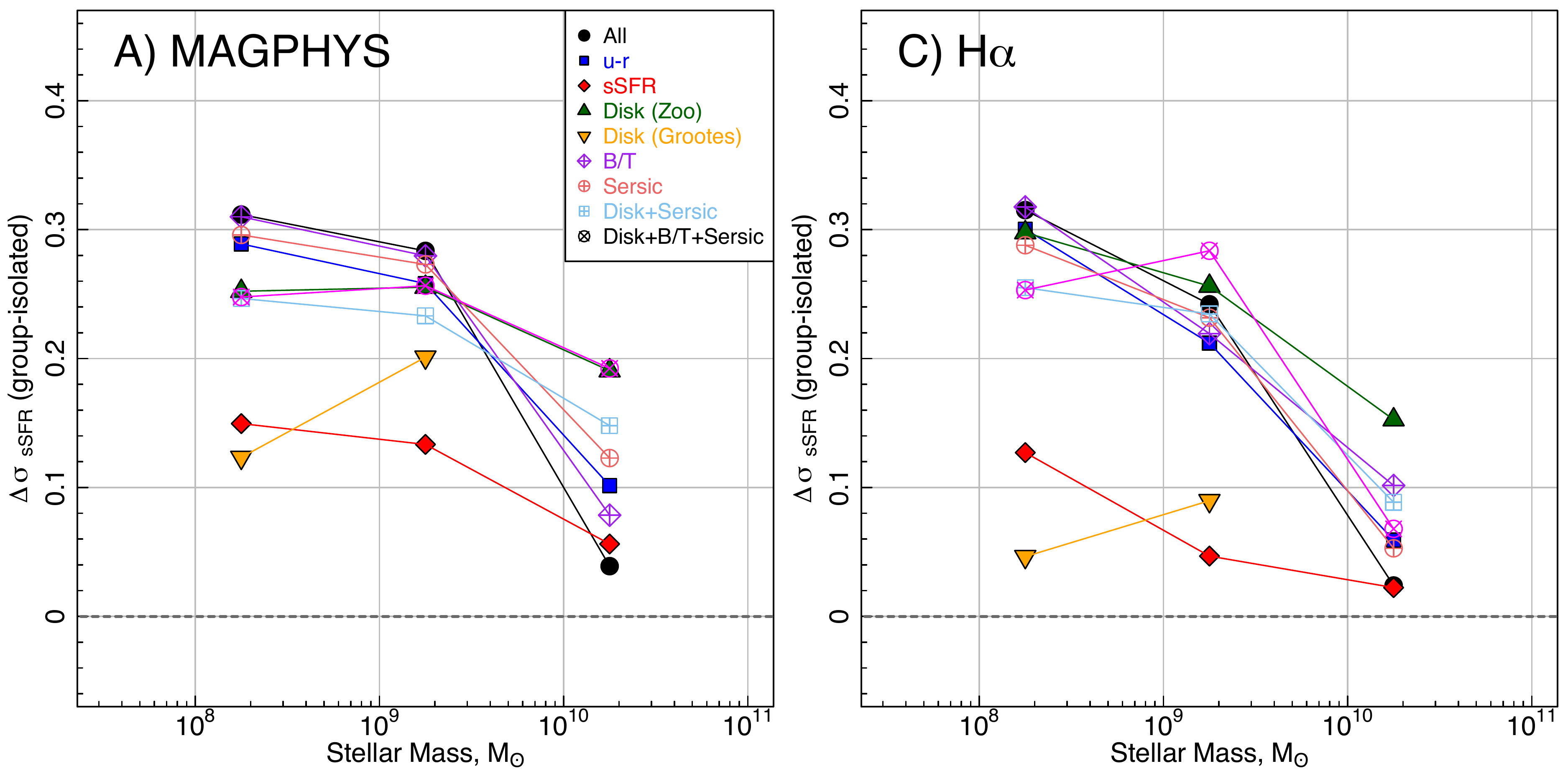}
\caption{Top: The same diagnostic displayed in Figure \ref{fig:slopes} but for just \textsc{magphys} and H$\alpha$-derived SFRs in group satellite galaxies. The differences between these panels and Figure \ref{fig:slopes} highlight additional sources of scatter in the $\sigma_{\mathrm{sSFR}}$-M$_{*}$ relation in groups (see text for details). Bottom: The relative difference between $\sigma_{\mathrm{sSFR}}$ for group satellites and $\sigma_{\mathrm{sSFR}}$ for isolated galaxies in three stellar mass bins. We find increased dispersion in group satellites at all stellar masses ($\Delta \sigma_{\mathrm{sSFR}}>0$), which is more extreme at low and intermediate masses.}
\label{fig:slopesGroup}
\end{center}
\end{figure*}

In order to easily compare the shape of the $\sigma_{\mathrm{sSFR}}$-M$_{*}$ relation for all SFR indicators and sample selections, we define a simple metric based on the slope of $\sigma_{\mathrm{sSFR}}$-M$_{*}$ in the low- and high-stellar mass regime; divided at log$_{10}$[M$_{*}$/M$_{\odot}$]=9.25. To choose this reference point we fit a 2$^{nd}$ order polynomial to the $\sigma_{\mathrm{sSFR}}$-M$_{*}$ relation for all samples and SFR indicators and find the median minimum point to be at log$_{10}$[M$_{*}$/M$_{\odot}$]=9.261, hence our closest stellar mass bin is log$_{10}$[M$_{*}$/M$_{\odot}$]=9.25. We then perform a least squares linear fit to the low mass and high mass  $\sigma_{\mathrm{sSFR}}$-M$_{*}$ respectively. This process is described visually in the bottom right panel of Figure \ref{fig:slopes}. While the point at log$_{10}$[M$_{*}$/M$_{\odot}$]=9.25 does not represent the minimum in all distributions (see above), we wish to use a consistent point across all $\sigma_{\mathrm{sSFR}}$-M$_{*}$ relations to highlight any variation with sample selection and/or SFR indicator. However, we do repeat this analysis for slopes about both log$_{10}$[M$_{*}$/M$_{\odot}$]=8.75 and log$_{10}$[M$_{*}$/M$_{\odot}$]=9.75 and find consistent results. While this is potentially a crude metric for the shape of the distributions, it allows an easy comparison of multipule different samples in the same figure. 

Figure \ref{fig:slopes} displays these low- and high-mass slopes for each sample selection (colours) and for each SFR indicator (panels). In this figure, points which span the $y=0$ line have a `U'-shaped distribution. In addition, the separation between the points (length of the coloured lines) describes how curved the distribution is, with points close together indicating a linear relation, and far apart a highly curved distribution. For samples that cross the $y=0$ line, the position where the coloured line crosses the $y=0$ line describes the asymmetry of the distribution about log$_{10}$[M$_{*}$/M$_{\odot}$]=9.25. For example, a purely symmetric `U'-shape would have widely-vertically-spaced points with the dashed $y=0$ crossing at the midpoint of their connecting line. 

From this figure we see that the majority of our samples have widely-separated points that span the y=0 line. This metric displays, at a glance, that irrespective of sample selection or SFR indicator the $\sigma_{\mathrm{sSFR}}$-M$_{*}$ relation is U'-shaped with minimum close to log$_{10}$[M$_{*}$/M$_{\odot}$]=9.25 and increasing dispersion to low and high stellar masses. The exception to this is the selection using sSFR which has a linearly decreasing dispersion with stellar mass in the majority of cases (both points sit below the  y=0 line). Once again this is expected as the cut in sSFR removes the large dispersion population at high stellar masses.

\subsection{Comparison To The State-Of-The-Art Semi Analytic Model and Hydrodynamical Simulations }
\label{sec:shark}

In addition to the Katsianis et al (in preparation) results from EAGLE discussed previously, we also compare our observed $\sigma_{\mathrm{sSFR}}$-M$_{*}$ relation to the state-of-the-art semi analytic model (SAM) Shark \citep{Lagos18}. Briefly, Shark is a highly flexible and modular open source SAM which includes several different models for gas cooling, active galactic nuclei, stellar and photo-ionisation feedback, and star formation. Here we use the Shark parameters described in \cite{Lagos18} run over the SURFS \citep{Elahi18} simulations with Planck15 cosmology and take the Shark+SURFS galaxies in the z=0 shapshot. We use both the medi-SURFS (medium box size and medium mass resolution) and mircro-SURFS (small box size and high mass resolution) simulation runs \citep[we do not go into further details of the simulations here, for more details see][]{Elahi18,Lagos18}. Using the Shark simulated galaxies initially we take all isolated central galaxies (as in GAMA) and bin in stellar mass bins of log$_{10}$[M$_{*}$/M$_{\odot}$]=0.2\,dex. We then calculate the standard deviation of sSFRs in each bin (consistent with our observational data). For details of the EAGLE results and in depth discussion of the EAGLE $\sigma_{\mathrm{sSFR}}$-M$_{*}$ relation, see Katsianis et al. 

The left panel of Figure \ref{fig:shark} displays the $\sigma_{\mathrm{sSFR}}$-M$_{*}$ relation using all SFR indicators used in this work for all isolated galaxies overlaid with the relations measured from both Shark and EAGLE (Note that for EAGLE we also only include all isolated central galaxies at z$\sim0$). We find that the $\sigma_{\mathrm{sSFR}}$-M$_{*}$ relation from both simulations are consistent with the GAMA results; showing a `U'-shape with minimum at log$_{10}$[M$_{*}$/M$_{\odot}$]=9-10 and a steep rise at higher stellar masses (albeit only over a small stellar mass range for EAGLE). We also highlight  the gradual rise to lower stellar masses in the Shark results, which is observed in some of our SFR indicators. This falls below the mass resolution limit for EAGLE. 

We also replicate the sample with a log$_{10}$[sSFR, yr$^{-1}$]$>$-11 cut (right panel of Figure \ref{fig:shark}), where both EAGLE and Shark follow a similar relation to the observational data at low to intermediate stellar masses, which sits within the spread of the different SFR indicators explored here. At the highest stellar masses we still see an upturn in $\sigma_{\mathrm{sSFR}}$ for Shark which is not found in EAGLE or the observations. This is due to the fact that in Shark, star formation in very massive galaxies declines slowly and galaxies stay above the sSFR cut. In EAGLE massive galaxies tend to quench rapidly, dropping below the sSFR cut. We also note that the EAGLE mass resolution is limited at log$_{10}$[M$_{*}$/M$_{\odot}$]$\sim$9 and the resolution limit for Shark is log$_{10}$[M$_{*}$/M$_{\odot}$]$\sim$8 for medi-SURFS and log$_{10}$[M$_{*}$/M$_{\odot}$]$\sim$7 for micro-SURFS.

In summary, the Shark model reproduces the observed $\sigma_{\mathrm{sSFR}}$-M$_{*}$ relation across all stellar masses incredibly well given that it was not specifically tuned to reproduce the dispersion in this plane, while EAGLE reproduces the relation well above its resolution limit. We note here again that within EAGLE, Katsianis et al attribute the high dispersion at low and high stellar masses to stellar and AGN feedback respectively. It is worth highlighting, that both EAGLE and Shark use a physical model for AGN and stellar feedback ($i.e.$ not a simple mass scaling which may artificially produce these results). In addition, \citep{Lagos18} showed that the scatter of the SFS is very sensitive to the adopted star formation law; that is both the way that the inter-stellar medium is partitioned between ionised, atomic and molecular gas, and the way molecular gas is converted into stars. Various theoretical and empirical models for star formation were used leading  to $>0.2 $\,dex differences in $\sigma_{\mathrm{sSFR}}$. These results suggest that the interplay between gas/stars within the galaxy can also play a significant role in shaping the $\sigma_{\mathrm{sSFR}}$-M$_{*}$ relation.

\subsection{The Impact of Group Environment}
\label{sec:group}

Finally, we consider the impact of group-scale environments on the metric used to parameterise the shape of the $\sigma_{\mathrm{sSFR}}$-M$_{*}$ relation in Section \ref{sec:var}.  Differences between these metrics for isolated and group environments can potentially elucidate additional physical sources of dispersion in the sSFR-M$_{*}$ relation caused by group astrophysical processes. To reduce confusion in the number of samples displayed, in this section we will only focus on \textsc{magphys} and H$\alpha$-derived SFRs.    

To explore the effect of group environment, we select all galaxies within our volume-limited samples which are in a $N>2$ group from the GAMA group catalogue of \cite{Robotham11}. We then exclude all sources which are closest to the iterative centre of the group as a central. The remaining sources form our `group satellites' sample, for which we repeat all of the analysis in the previous sections. 

The top row of Figure \ref{fig:slopesGroup} displays the same metric as in Figure \ref{fig:slopes} but for the group satellites. There are two notable differences: i) the pairs of points for each sample selection are closer together indicating that the distribution is less curved, and ii) the points are more negative both at the high (squares) and low (circles) mass end indicating that in most cases the distributions are no longer `U'-shaped but have a decreasing $\sigma_{\mathrm{sSFR}}$ with increasing stellar mass. 

\begin{figure}
\begin{center}
\includegraphics[scale=0.28]{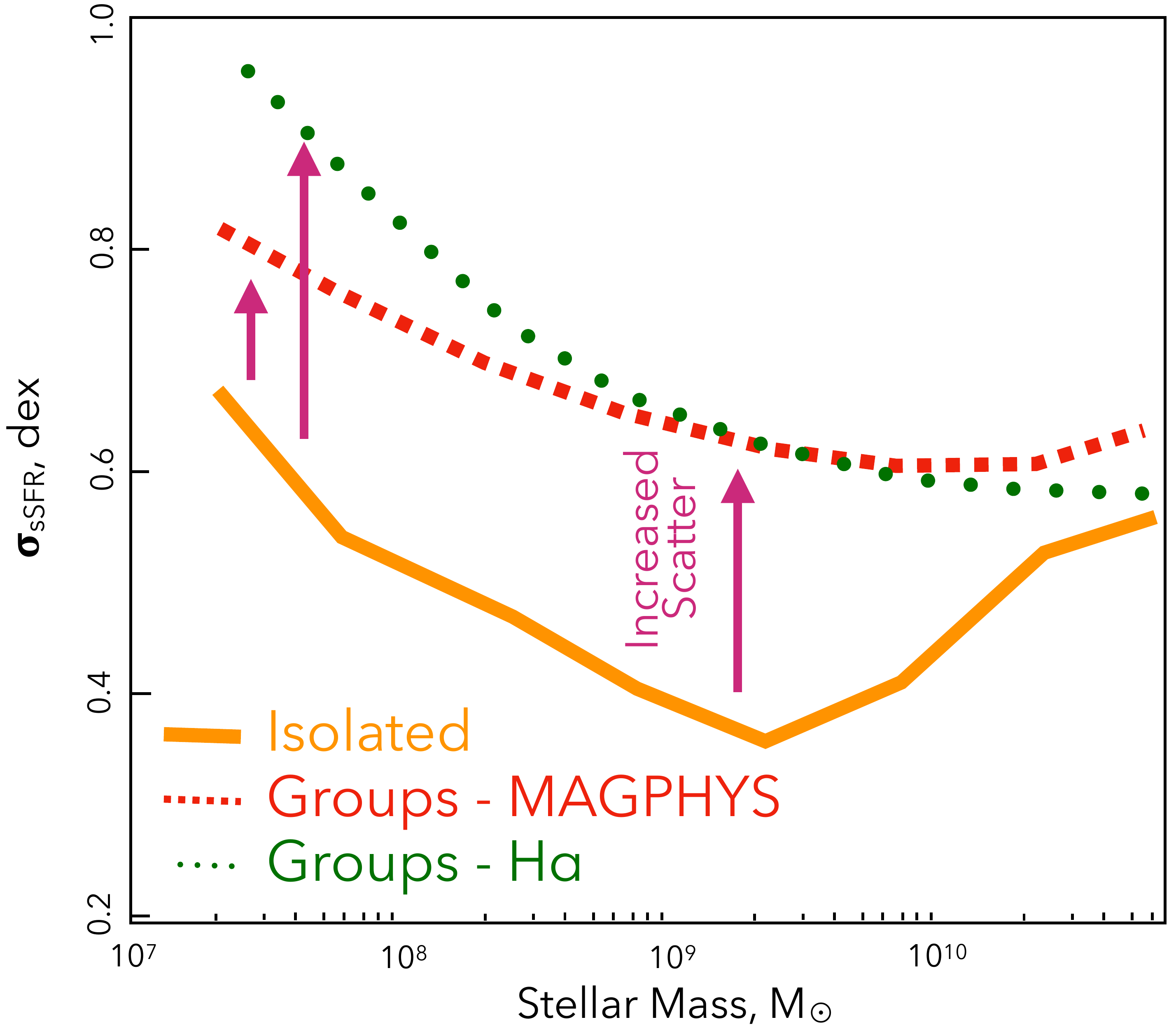}
\caption{Cartoon representation of the differences in observed $\sigma_{\mathrm{sSFR}}$-M$_{*}$ relation between isolated and group environments. The parabolic distribution observed in isolated galaxies is removed by additional large dispersion at low and intermediate stellar masses.}
\label{fig:cartoon}
\end{center}
\end{figure}

However, the top row of Figure \ref{fig:slopesGroup} does not inform as to whether the observed changes are caused by the dispersion increasing or decreasing in group environments, only that the relative shape of the $\sigma_{\mathrm{sSFR}}$-M$_{*}$ relation changes. To explore this further, the bottom row of Figure \ref{fig:slopesGroup} displays the normalisation difference of $\sigma_{\mathrm{sSFR}}$ between the isolated and group satellite samples at three stellar mass bins  ($i.e.$ for each independent sample selection and SFR indicator we measure the offset between log$_{10}$[$\sigma_{\mathrm{sSFR}}$] in isolated and group environments at log$_{10}$[M$_{*}$/M$_{\odot}$]$=$8.25, 9.25 and 10.25). We find that $\sigma_{\mathrm{sSFR}}$ is larger in group satellites than in isolated galaxies for all samples and at all stellar masses ($i.e.$ all points are above zero), but that the increase in dispersion is much larger at low and intermediate stellar masses than at high stellar masses. This indicates that group environments preferentially increase $\sigma_{\mathrm{sSFR}}$ for low mass galaxies, likely due to increased satellite quenching of low mass galaxies leading to increased dispersion (this is explored in detail in Davies et al, in preparation).  To highlight this, Figure \ref{fig:cartoon} shows a cartoon representation of how the $\sigma_{\mathrm{sSFR}}$-M$_{*}$ relation changes between isolated and group satellite galaxies. 

\begin{figure*}
\begin{center}
\includegraphics[scale=0.5]{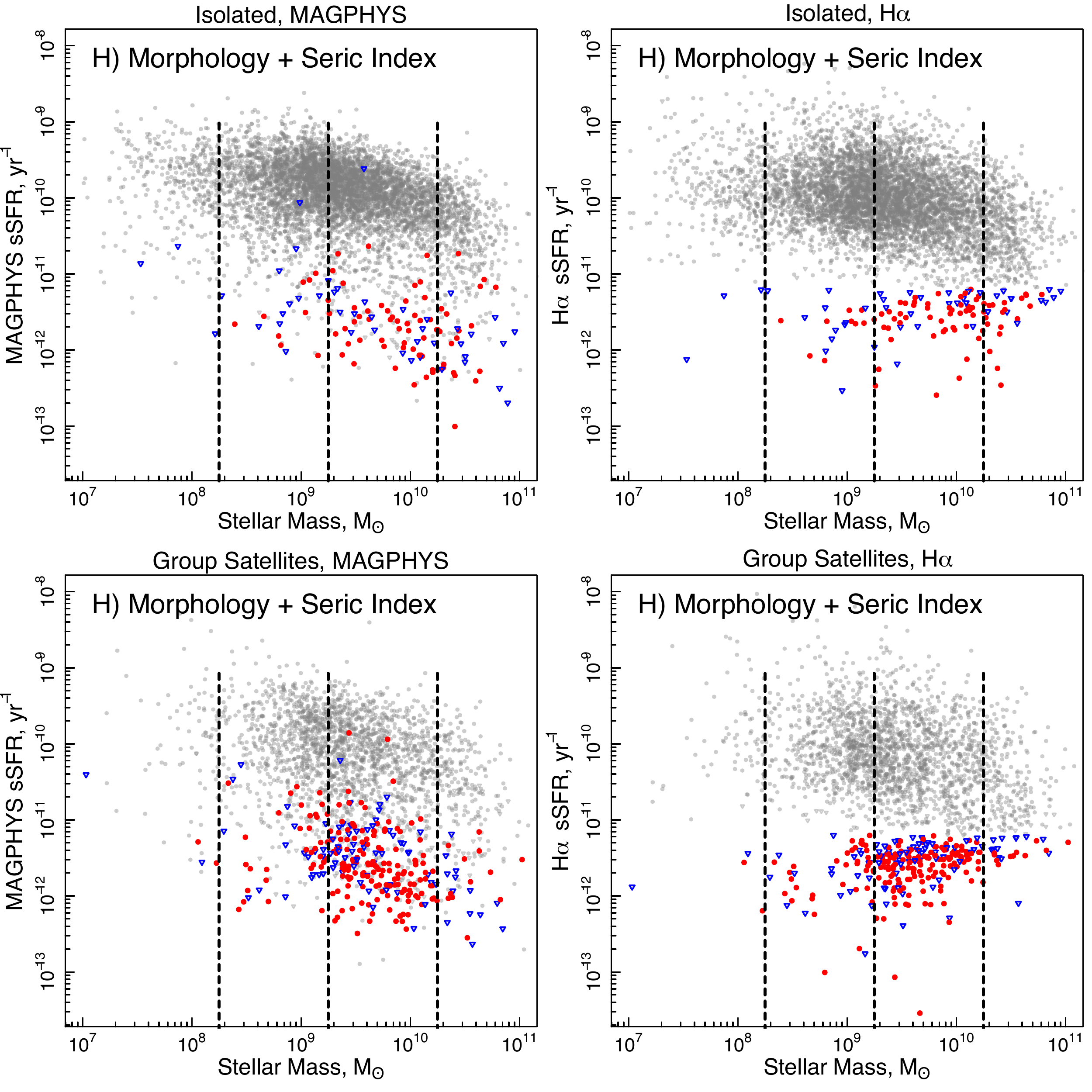}
\caption{The sSFR-M$_{*}$ plane for sources selected via Galaxy Zoo morphology and S\'{e}rsic index for isolated galaxies (top row) and group satellites (bottom row). H$\alpha$ and \textsc{magphys} SFRs are shown in the left and right column respectively. Dashed vertical lines display the points at which we measure the normalisation difference in $\sigma_{\mathrm{sSFR}}$ for Figure \ref{fig:slopesGroup}. Circles display source which have H$\alpha$ detected at signal-to-noise$>2$, while triangles display sources with H$\alpha$ signal-to-noise$<=2$. To highlight the additional quenched groups satellite population sources with sSFR$_{\mathrm{H}\alpha}<10^{11.2}$ are shown in colour (red circles and blue triangles). }
\label{fig:interesting}
\end{center}
\end{figure*}

To explore the populations which are driving the observed differences between isolated galaxies and group satellites further, we return to the full sSFR-M$_{*}$ plane and compare the distribution of sources in each environment. Figure \ref{fig:interesting} displays the H$\alpha$ and \textsc{magphys} sSFR-M$_{*}$ plane for isolated galaxies (top rows) and group satellites (bottom row) selected as disc-like systems using the combined Galaxy Zoo morphology and S\'{e}rsic index selection. The dashed vertical lines display the points at which we measure the normalisation difference in $\sigma_{\mathrm{sSFR}}$ for Figure \ref{fig:slopesGroup}.

Firstly, we find that there is uniformly larger dispersion along the SFS population at all stellar masses in groups; potentially due to more stochastic star formation processes occurring in groups via interactions. We also find an additional population of quenched group satellite galaxies at log$_{10}$[M$_{*}$/M$_{\odot}$]$\sim$9-10. To highlight this, in Figure \ref{fig:interesting} we colour all galaxies with sSFR$_{H\alpha}<10^{-11.2}$ in red. This population increases $\sigma_{\mathrm{sSFR}}$ at the point where  $\sigma_{\mathrm{sSFR}}$-M$_{*}$ is a minimum in isolated galaxies, removing the `U'-shape shape. This population is likely produced by group quenching processes (such as strangulation/starvation/stripping). We leave detailed discussion of potential physical mechanisms which may produce these populations to the following papers in this series. 

However, interestingly we note that this population sits on a tight sequence when using H$\alpha$-derived SFRs, but is much more dispersed to higher sSFR when using \textsc{magphys}-derived SFRs. This potentially indicates that group satellites which appear passive in H$\alpha$-derived SFRs are still transitioning across the green valley when using \textsc{magphys}-derived SFRs. There are two important differences between SFRs measured using these indicators: i) GAMA H$\alpha$ SFRs are derived from fibre-fed spectroscopy and therefore only probe the central regions of galaxies while \textsc{magphys} SFRs are integrated over the whole galaxy, and ii) H$\alpha$ SFRs probe a much shorter integrated physical timescale than \textsc{magphys} SFRs \citep[$i.e.$ $<$20\,Myr as compared to $<$100\,Myr. $e.g.$][]{Davies15a}. The differences in the red points in different columns of Figure \ref{fig:interesting} could be attributed to sources which have quenched on short timescales, and thus still have residual star-formation when measured using \textsc{magphys}, or sources which are undergoing inside-out quenching and thus become completely passive over the central regions observed by our fibre-fed spectroscopy prior to the whole galaxy becoming passive. We leave detailed discussion of these mechanisms to the following papers where we aim to produce a more complete picture of the sSFR-M$_{*}$ plane based on multipule observations.

\section{Conclusions}
\label{sec:conclusions}

In the first paper of this series we have parameterised the dispersion along the star-forming sequence for different sample selections and SFR indicators within GAMA. We find that:
\\
\\
$\bullet$ Irrespective of selection method of star-forming disc-like systems the $\sigma_{\mathrm{sSFR}}$-M$_{*}$ relation is parabolic with high dispersion at the low and high stellar mass end and minimum point of $\sigma_{\mathrm{sSFR}}\sim0.35-0.5$\,dex at log$_{10}$[M$_{*}$/M$_{\odot}$]$\sim$8.75-9.75. This holds true for different SFR indicators and when using mixture modelling, albeit with differences in normalisation of $\sigma_{\mathrm{sSFR}}$. \\      
\\
$\bullet$ Our results are largely consistent with a number of previous studies and simulation results from EAGLE, Illustris and Shark. The EAGLE study suggests that this parabolic dispersion is produced via stellar and AGN feedback at the low and high mass end respectively, while Shark suggests that the interplay between gas/stars within the galaxy can also play a significant role in shaping the dispersion. This will be explored further observationally in subsequent papers in this series. \\
\\
$\bullet$ In combination, these results suggests that the `U'-shape of $\sigma_{\mathrm{sSFR}}$-M$_{*}$ relation is physical and not an artefact of sample selection or method.\\
\\
\\
$\bullet$ For group satellite galaxies we find an increased dispersion at all stellar masses potentially due to more stochastic star formation processes occurring in groups. We also find an additional population of quenched sources at log$_{10}$[M$_{*}$/M$_{\odot}$]$\sim$9-10 which increases $\sigma_{\mathrm{sSFR}}$ and removes the parabolic shape. \\
\\
$\bullet$ We tentatively suggest that this population is produced by group quenching processes and highlight that differences between H$\alpha$- and \textsc{magphys}-derived SFRs for this population may be indicative of short timescale quenching. However, we leave detailed discussion to the following papers.  
\\

We have parametrised the $\sigma_{\mathrm{sSFR}}$-M$_{*}$ relation in the local Universe and highlighted that it is a useful tool in exploring the recent SFH of galaxies both in term of the variation of dispersion as a function of stellar mass and in the differences between isolated galaxies and group satellites. In the following papers in this series we will explore the morphological evolution across the sSFR-M$_{*}$ plane and the physical processes which drive the formation of this fundamental relation.

\section*{Acknowledgements}

GAMA is a joint European-Australasian project based around a spectroscopic campaign using the Anglo- Australian Telescope. The GAMA input catalogue is based on data taken from the Sloan Digital Sky Survey and the UKIRT Infrared Deep Sky Survey. Complementary imaging of the GAMA regions is being obtained by a number of in- dependent survey programs including GALEX MIS, VST KiDS, VISTA VIKING, WISE, Herschel-ATLAS, GMRT and ASKAP providing UV to radio coverage. GAMA is funded by the STFC (UK), the ARC (Australia), the AAO, and the participating institutions. The GAMA website is \url{http://www.gama-survey.org/}.

We acknowledge the Virgo Consortium for making their simulation data available. The EAGLE simulations were performed using the DiRAC-2 facility at Durham, managed by the ICC, and the PRACE facility Curie based in France at TGCC, CEA, Bruyeres-le-Chatel.

CL has received funding from a Discovery Early Career Researcher Award (DE150100618) and by the ARC Centre of
Excellence for All Sky Astrophysics in 3 Dimensions (ASTRO 3D), through project number CE170100013.

\appendix

\section{Other SFR indicators}

\begin{figure*}
\begin{center}
\includegraphics[scale=0.47]{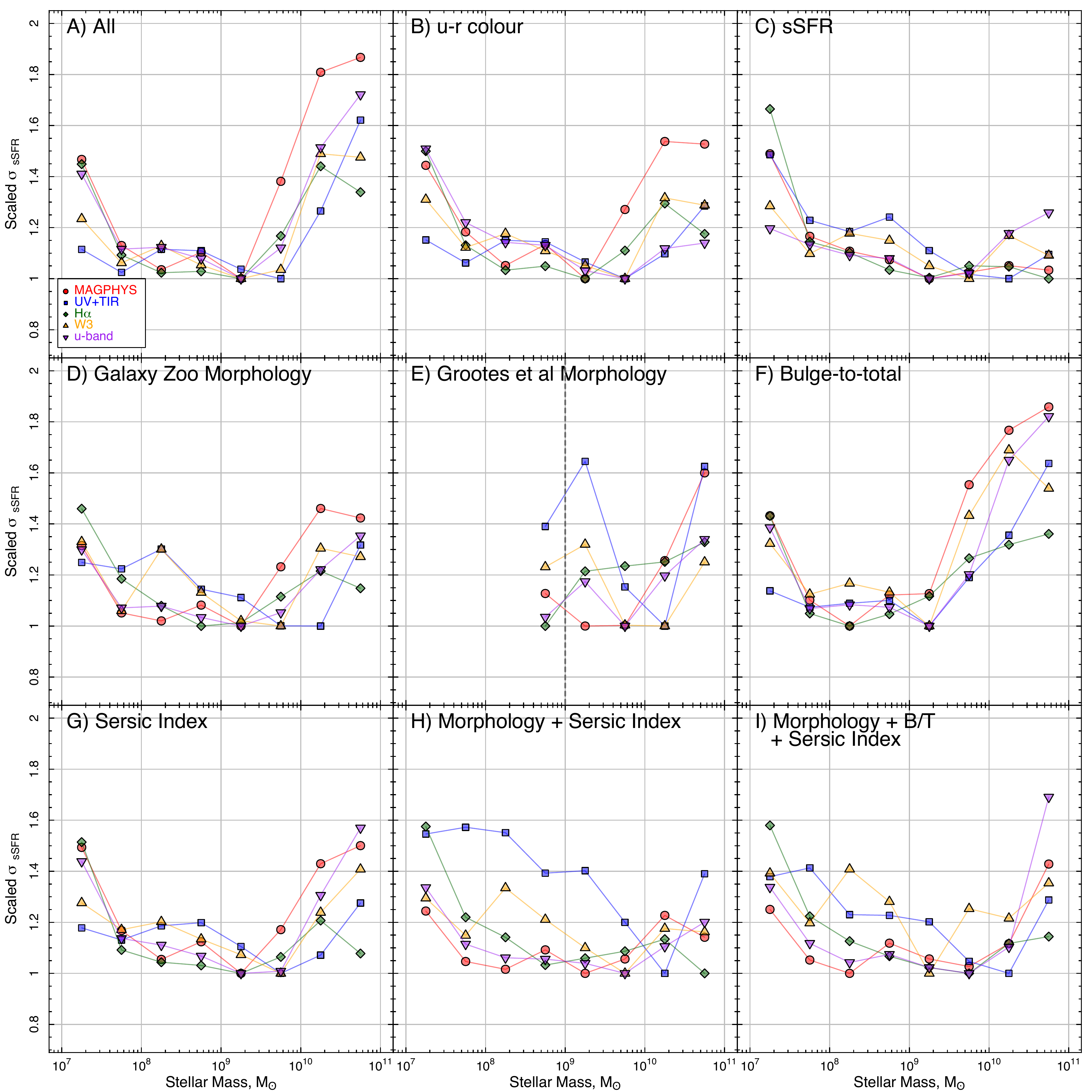}
\caption{The $\sigma_{\mathrm{sSFR}}$-M$_{*}$ relation for different SFR indicators and using each method for isolating the SFS. Figure follows the same layout as Figure \ref{fig:scatter}, but shows all SFR indicators in each panel - individual SFR indicator figures are given in the appendix. All distributions are scaled, dividing each line by the minimum dispersion for the given SFR indicator. This is intended to remove any dependancy on the absolute value of the dispersion and to  highlight variations in shape. The majority of the distributions show a `U-shaped'  $\sigma_{\mathrm{sSFR}}$-M$_{*}$ relation with minimum at $\sim10^{9.25}$\,M$_{\odot}$ and increased scatter at low and high masses. }
\label{fig:scatAll}
\end{center}
\end{figure*}

This appendix displays identical figures to Figure \ref{fig:selections} and \ref{fig:scatter} but for all other SFR indicators discussed in Section \ref{sec:SFRindicators}. Summaries of these figures are displayed in Figure \ref{fig:scatAll}. This figure displays the relation for all indicators, scaled to the log$_{10}$[M$_{*}$/M$_{\odot}$]=9.25 stellar mass bin. To choose this reference point we fit a 2$^{nd}$ order polynomial to the $\sigma_{\mathrm{sSFR}}$-M$_{*}$ relation for all samples and SFR indicators and find the median minimum point to be at log$_{10}$[M$_{*}$/M$_{\odot}$]=9.261, hence our closest stellar mass bin is log$_{10}$[M$_{*}$/M$_{\odot}$]=9.25. Here we can more clearly see that in almost all cases the  $\sigma_{\mathrm{sSFR}}$-M$_{*}$ relation follows a `U'--shaped distribution. We do find that the stellar mass of the minimum point varies in some cases.

\begin{figure*}
\begin{center}
\includegraphics[scale=0.31]{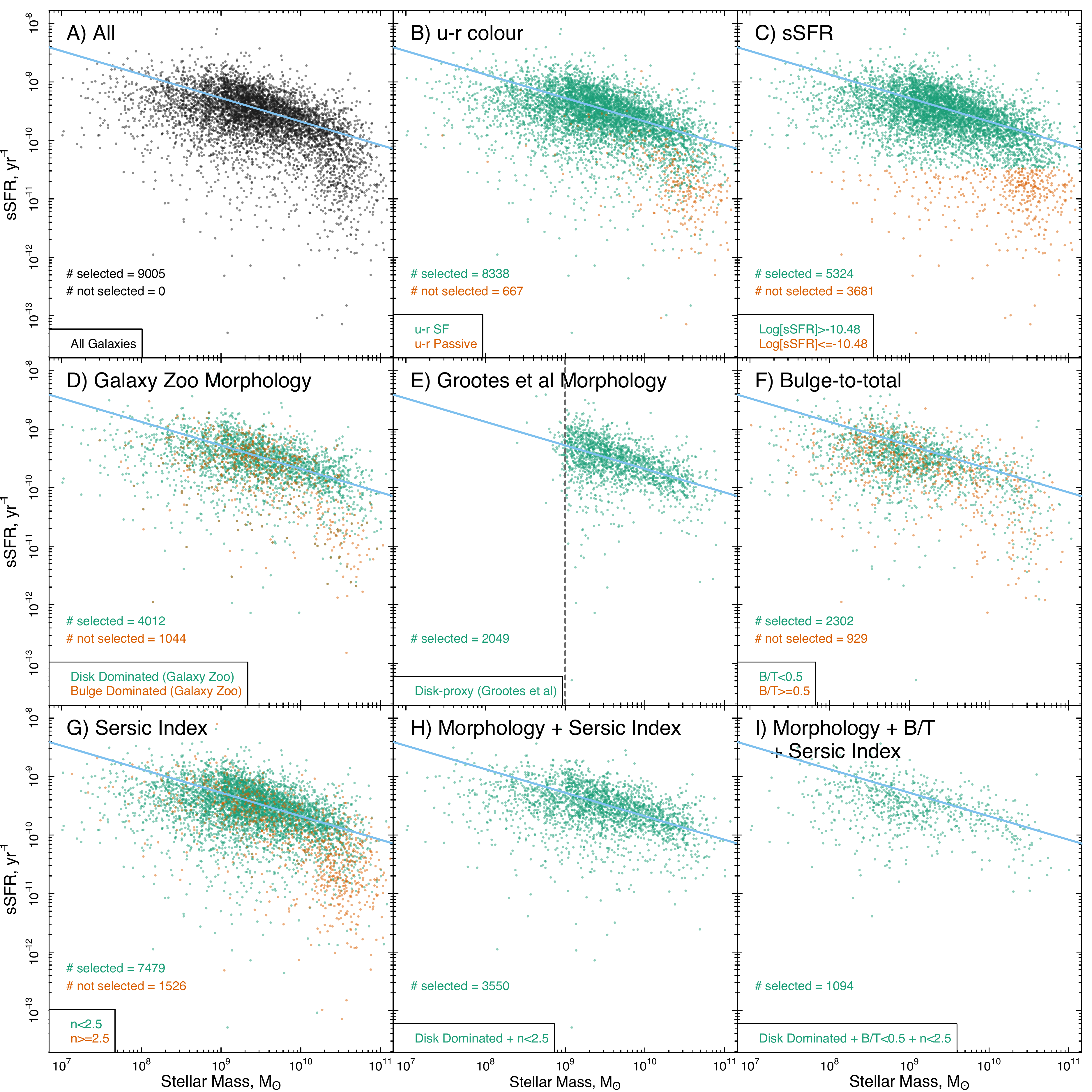}
\\

\includegraphics[scale=0.31]{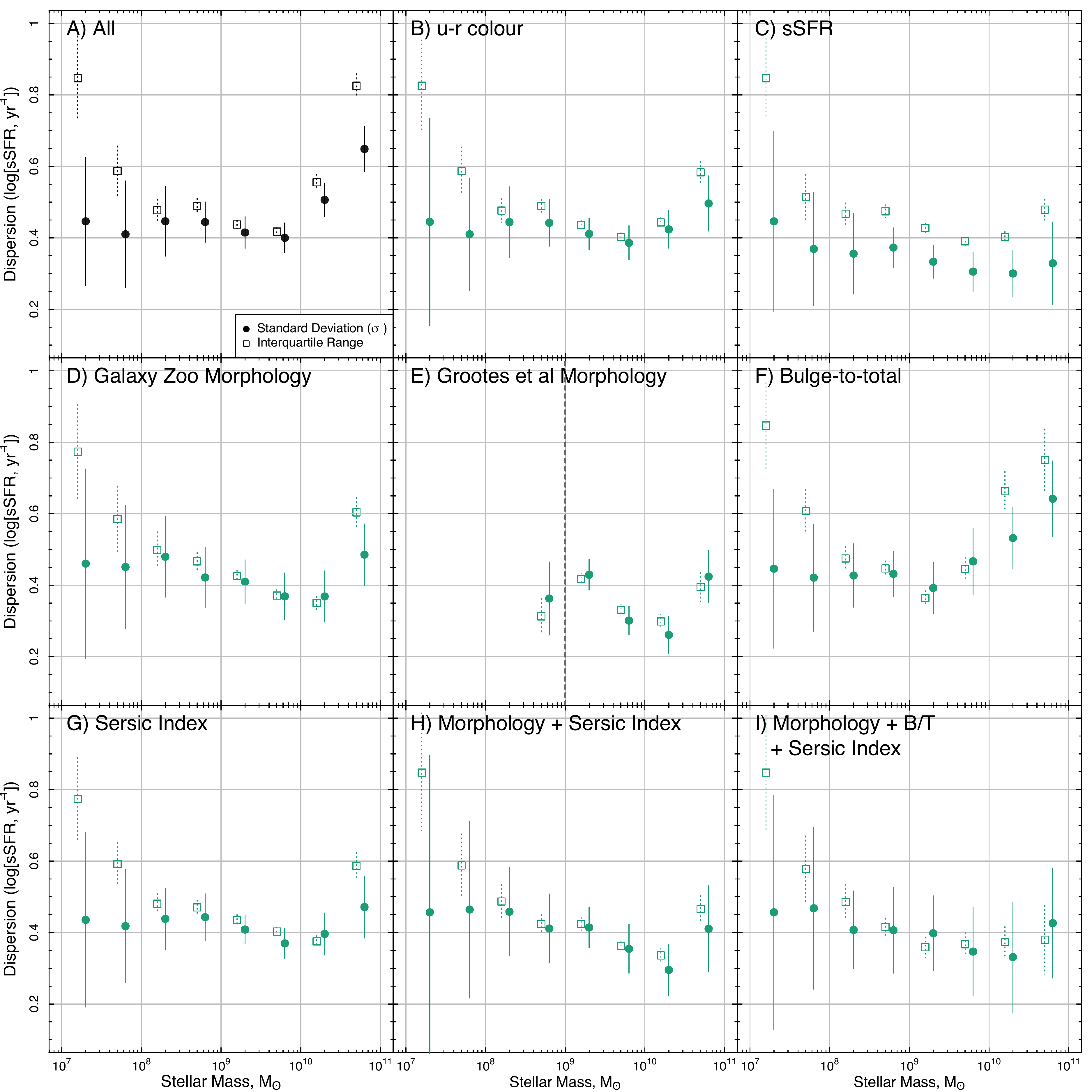}

\caption{Same as Figures 3 and 5 but for UV+TIR-derived SFRs}
\label{fig:scatterUVTIR}
\end{center}
\end{figure*}

\begin{figure*}
\begin{center}
\includegraphics[scale=0.31]{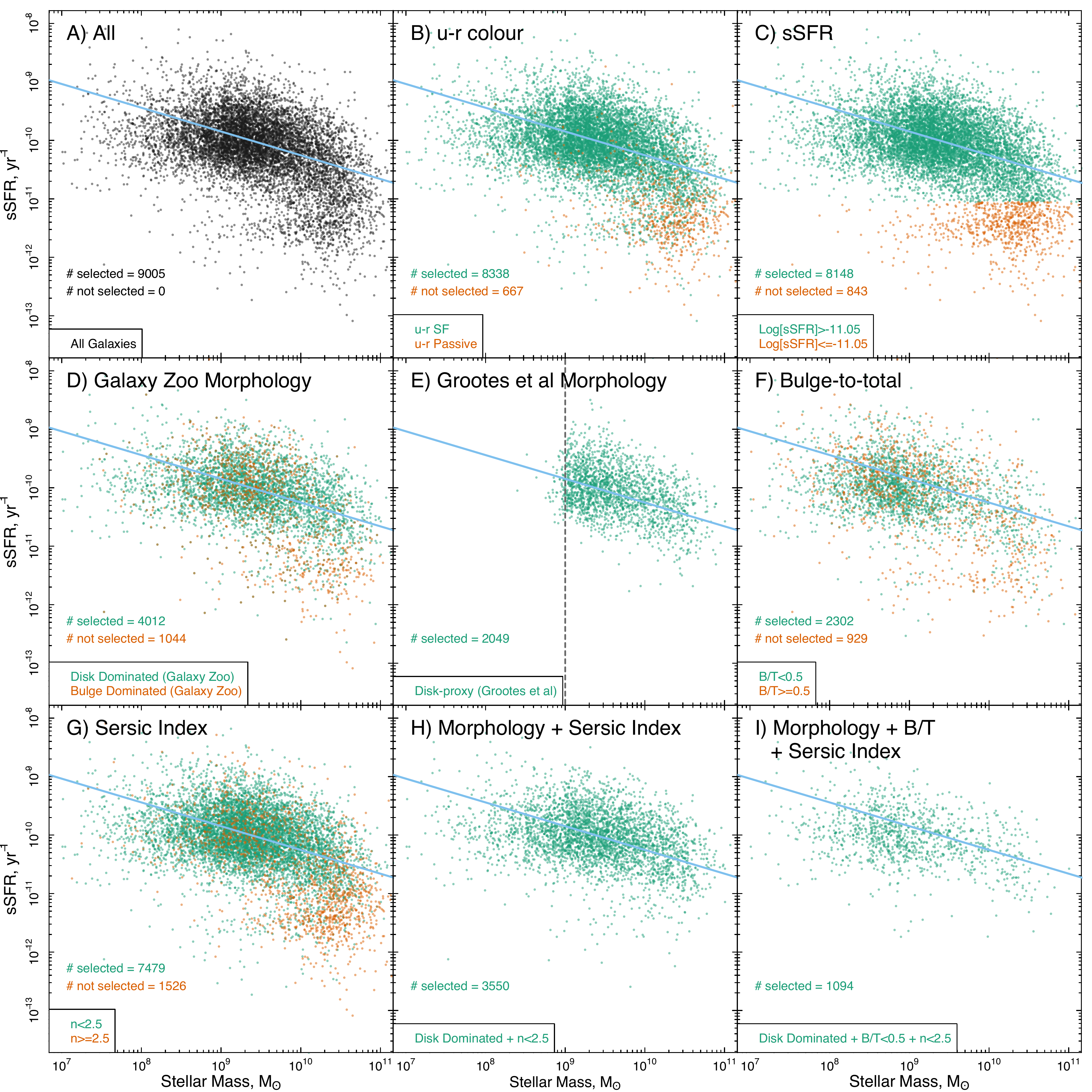}
\\

\includegraphics[scale=0.31]{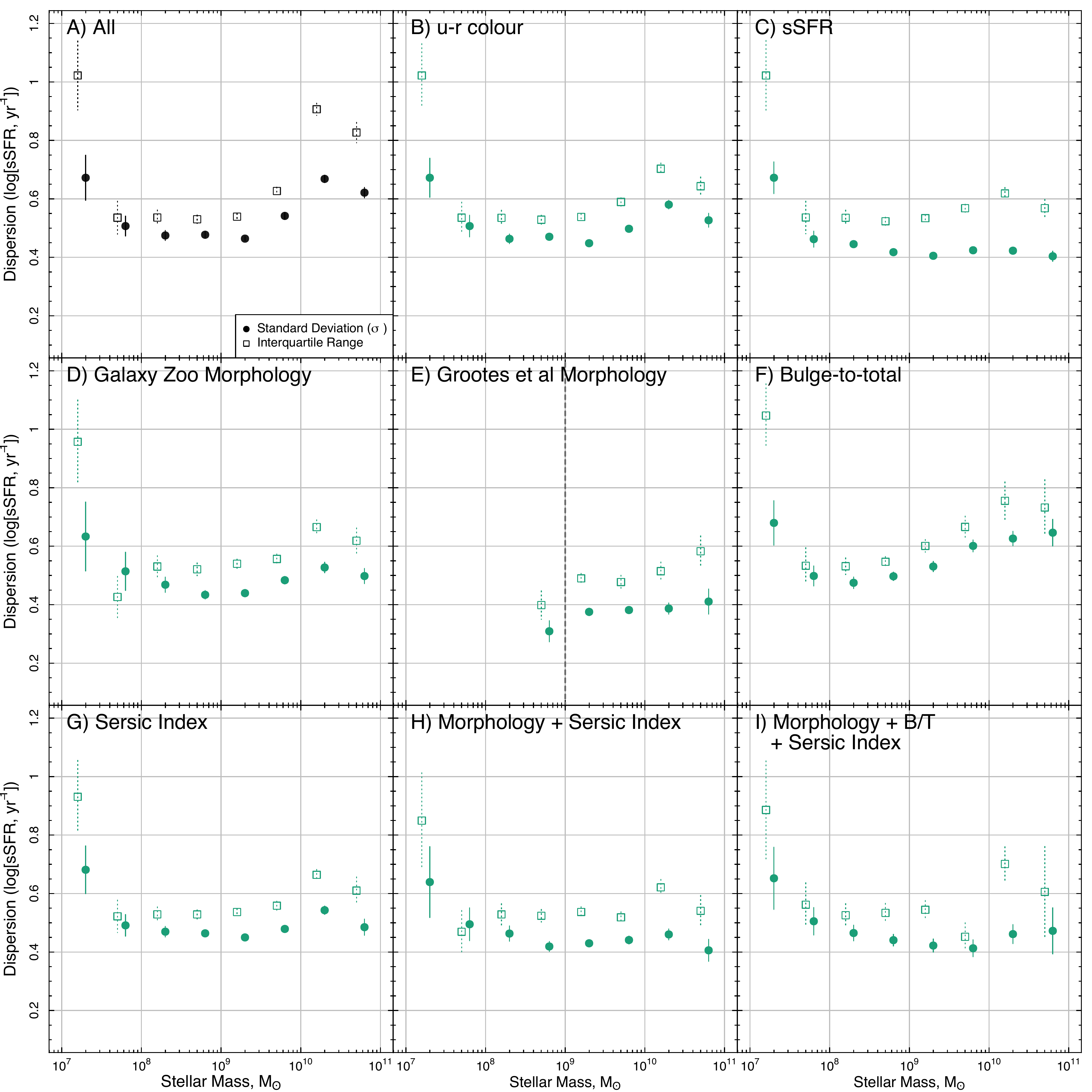}

\caption{Same as Figures 3 and 5 but for H$\alpha$-derived SFRs}
\label{fig:scatterHa}
\end{center}
\end{figure*}

\begin{figure*}
\begin{center}
\includegraphics[scale=0.31]{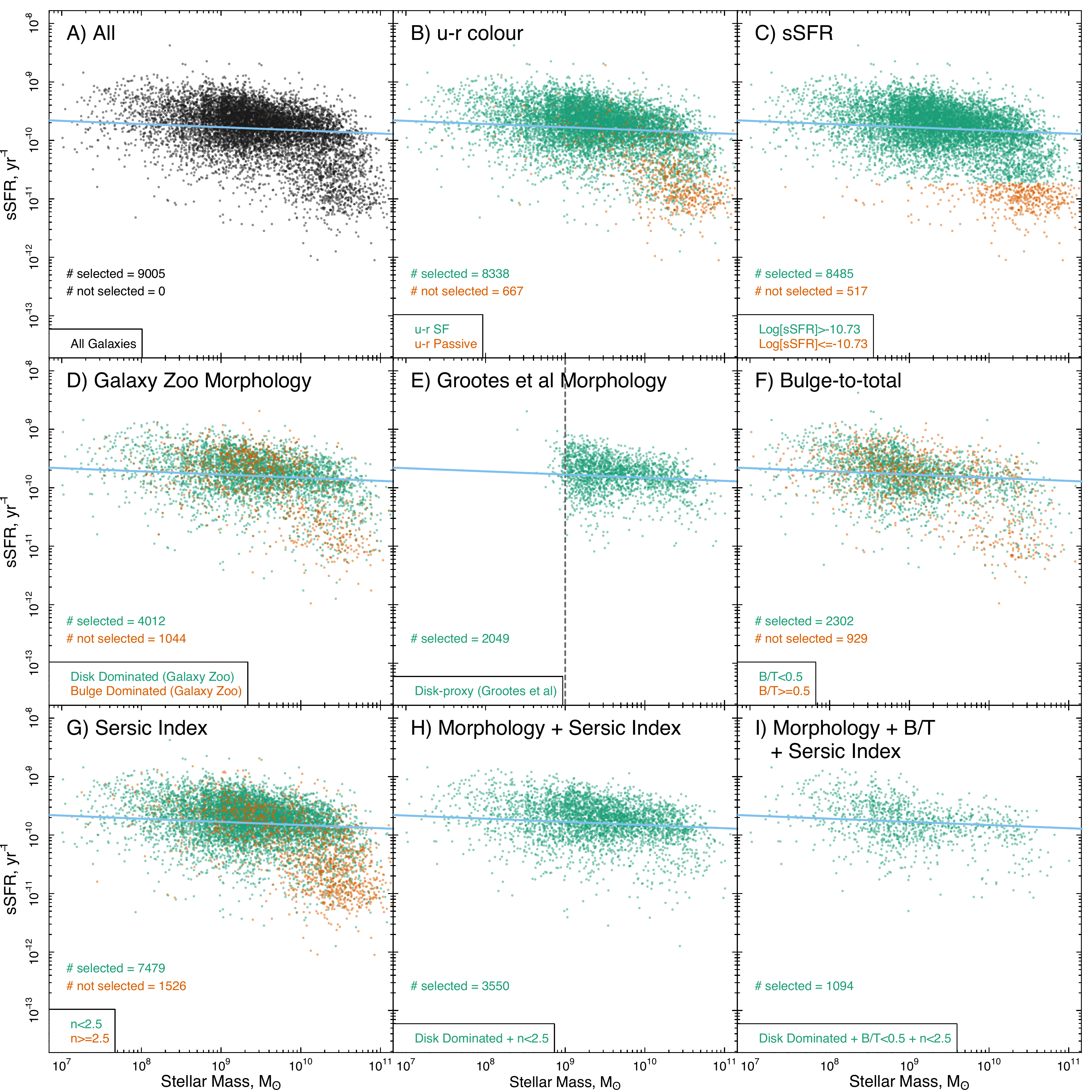}
\\

\includegraphics[scale=0.31]{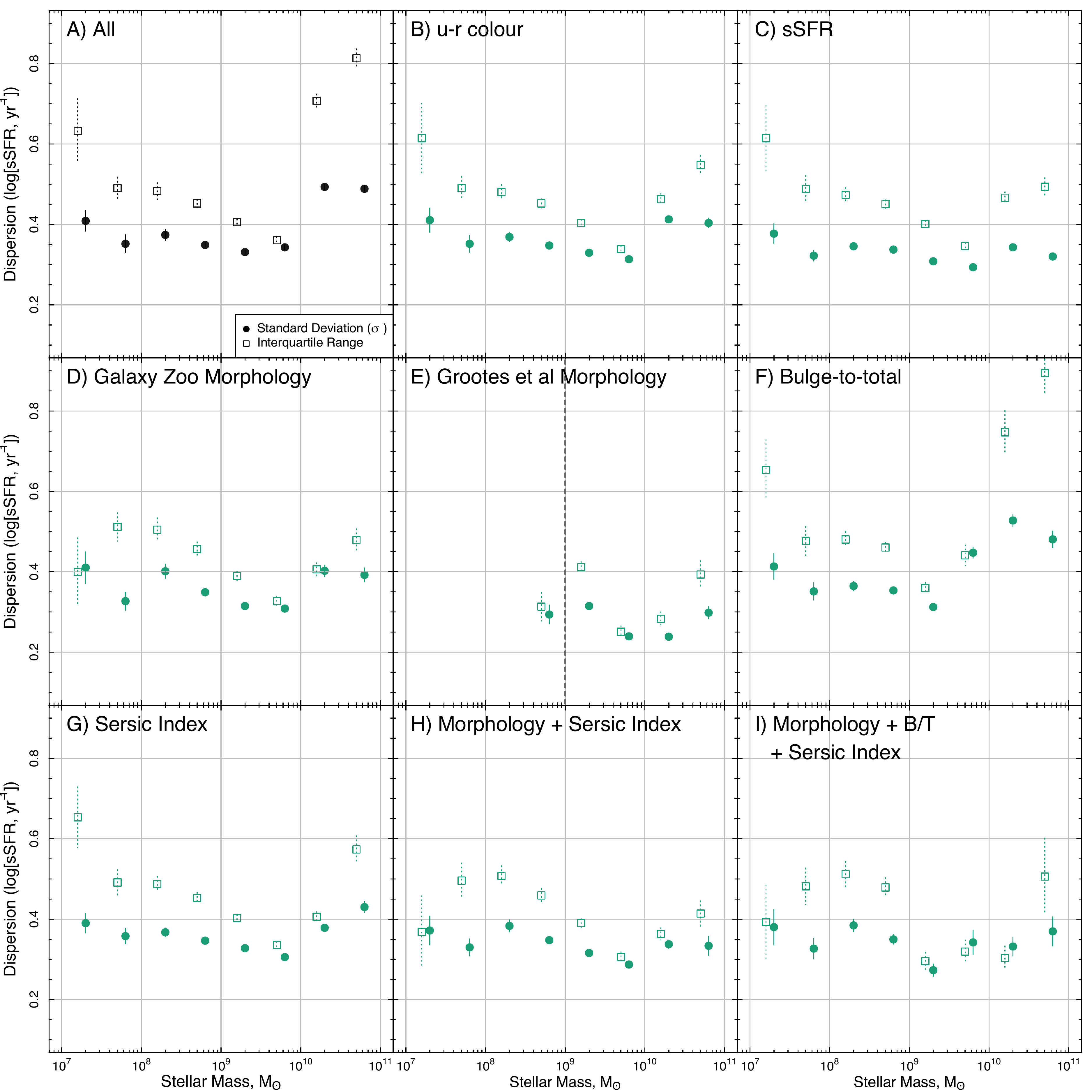}

\caption{Same as Figures 3 and 5 but for W3-derived SFRs}
\label{fig:scatterW3}
\end{center}
\end{figure*}

\begin{figure*}
\begin{center}
\includegraphics[scale=0.31]{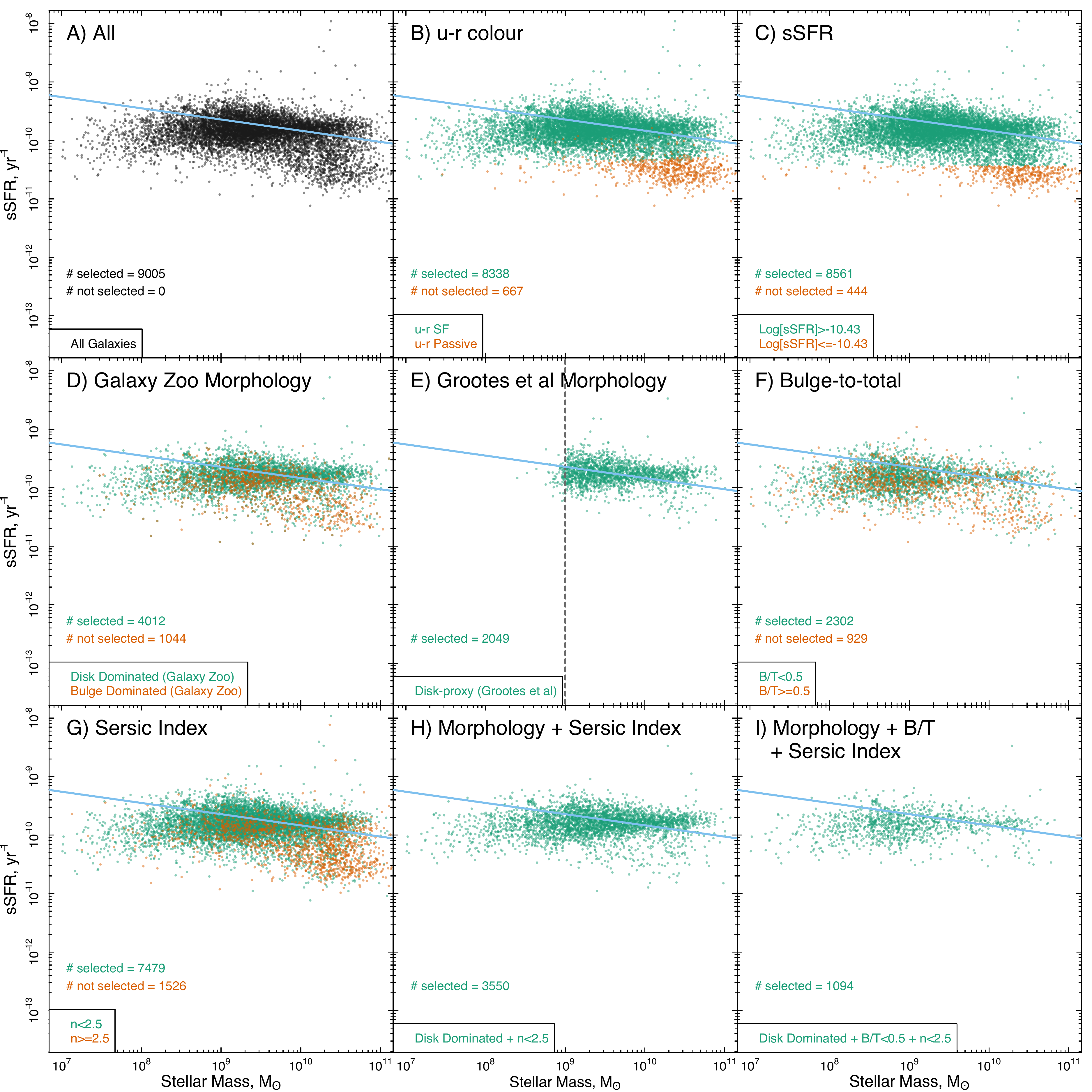}
\\

\includegraphics[scale=0.31]{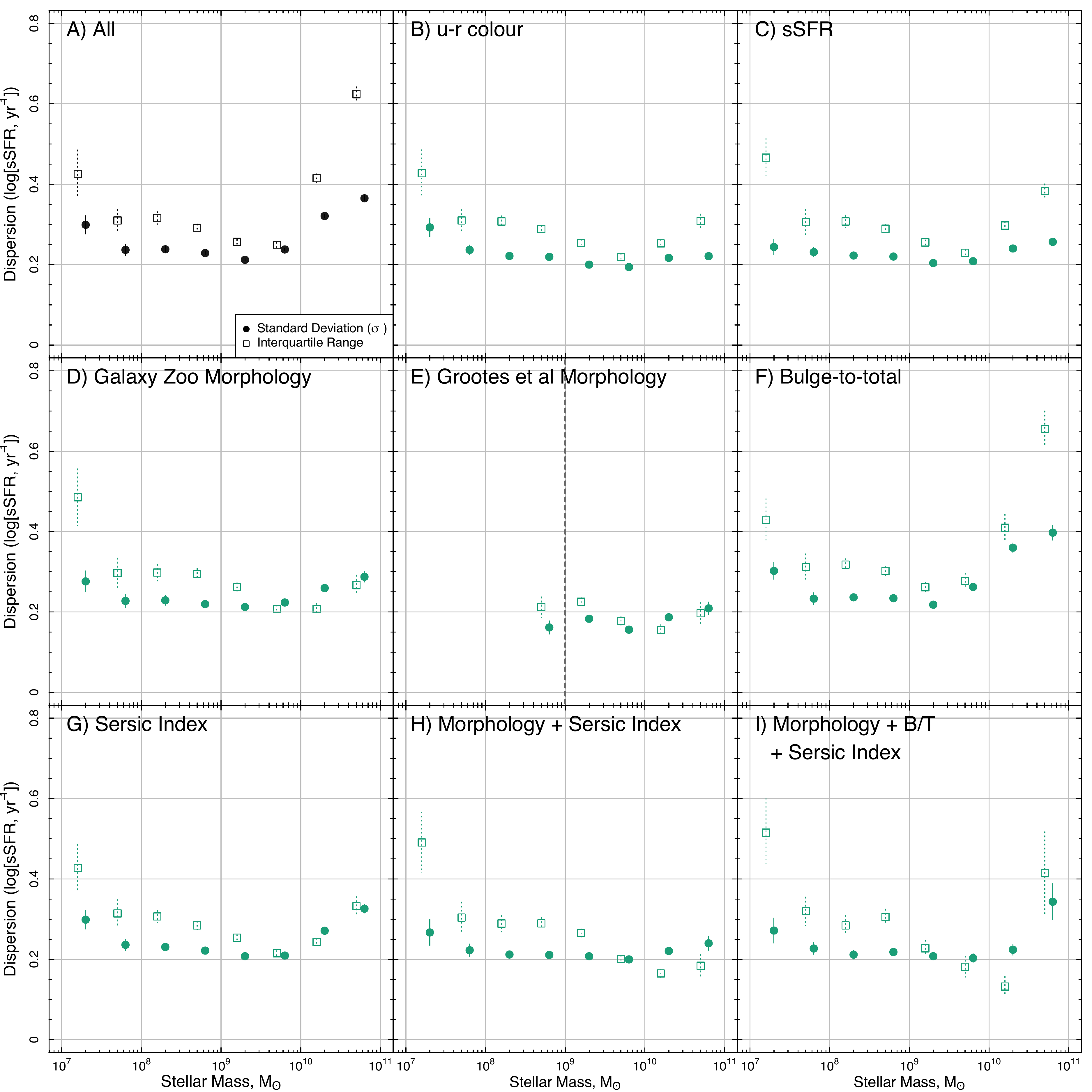}

\caption{Same as Figures 3 and 5 but for u-band-derived SFRs}
\label{fig:scatterU}
\end{center}
\end{figure*}

\bsp	
\label{lastpage}
\end{document}